\documentclass[12pt]{emulateapj}

\newcommand{\beq}{\begin{equation}}
\newcommand{\eeq}{\end{equation}}
\newcommand{\etal}{{\sl et~al.~}}
\newcommand{\kms}{km s$^{-1}$~}
\newcommand{\lC}{{$\ell$} Car}
\newcommand{\HST}{{\it HST}}
\newcommand{\HIP}{{\it HIPPARCOS}}

\def\fdg{\hbox{$.\!\!^\circ$}}
\def\arcmin{\hbox{$^\prime$}}
\def\lCpi{$ 2.01\pm0.20$~}
\def\zGpi{$ 2.78\pm0.18$~}
\def\bDpi{$ 3.14\pm0.16$~}

\def\WSpi{$ 2.28\pm0.20$~}
\def\XSpi{$ 3.00\pm0.18$~}
\def\YSpi{$ 2.13\pm0.29$~}
\def\FApi{$ 2.81\pm0.18$~}
\def\TVpi{$ 1.90\pm0.23$~}
\def\RApi{$ 2.40\pm0.19$~}
\def\dCpi{$ 3.66\pm0.15$~}

\received{}
\revised{}
\accepted{}

\shorttitle{Galactic Cepheid Period-Luminosity Relations}
\shortauthors{Benedict \etal}

\begin{document}
\bibliographystyle{plainnat}
\title{{\it Hubble Space Telescope} Fine Guidance Sensor Parallaxes of Galactic Cepheid Variable Stars: Period-Luminosity Relations\footnote{Based on 
observations made with
the NASA/ESA Hubble Space Telescope, obtained at the Space Telescope
Science Institute, which is operated by the
Association of Universities for Research in Astronomy, Inc., under NASA
contract NAS5-26555} }

\author{ G.\ Fritz  Benedict\altaffilmark{2}, Barbara E.
McArthur\altaffilmark{2}, Michael W. Feast\altaffilmark{3}, Thomas G. Barnes\altaffilmark{7}, Thomas E. Harrison\altaffilmark{4},\altaffilmark{8},  Richard J. Patterson\altaffilmark{5}, John W. Menzies\altaffilmark{9}, Jacob L. Bean\altaffilmark{2}, Wendy L. Freedman\altaffilmark{6}}

\altaffiltext{2}{McDonald Observatory, University of Texas, Austin, TX 78712}
\altaffiltext{3}{Astronomy Dept, University of Cape Town, Rondebosch, South Africa, 7701}
\altaffiltext{4}{Department of Astronomy, New Mexico State University, Las Cruces, NM 88003}
\altaffiltext{5}{Department of Astronomy, University of Virginia, Charlottesville, VA 22903}
\altaffiltext{6}{The Observatories, Carnegie Institution of Washington, Pasadena, CA
91101}
\altaffiltext{7}{McDonald Observatory, University of Texas, Austin, TX 78712; currently on assignment to the National Science Foundation, 4201 Wilson Boulevard, Arlington, VA 
22230.}
\altaffiltext{8}{Visiting Astronomer, Kitt Peak National Observatory, National
Optical Astronomy Observatory, which is operated by the Association of
Universities for Research in Astronomy, Inc., under cooperative agreement with
the National Science Foundation.}
\altaffiltext{9}{South African Astronomical Observatory, Observatory, South Africa, 7935}


\begin{abstract}
We present new absolute trigonometric parallaxes and relative proper motions for nine Galactic Cepheid variable stars: \lC, $\zeta$ Gem, $\beta$ Dor, W Sgr,  X Sgr, Y Sgr, FF Aql, T Vul, and RT Aur. We obtain these results with astrometric data from Fine Guidance Sensor 1r, a white-light interferometer on {\it Hubble Space Telescope}. We find absolute parallaxes  in milliseconds of arc:  \lC, \lCpi; $\zeta$ Gem, \zGpi; $\beta$ Dor, \bDpi; W Sgr, \WSpi;  X Sgr, \XSpi; Y Sgr, \YSpi; FF Aql, \FApi; T Vul, \TVpi; and RT Aur, \RApi, an average $\sigma_\pi/\pi$ = 8\%. Two stars (FF Aql and W Sgr) required the inclusion of binary astrometric perturbations, providing Cepheid mass estimates.
With these parallaxes we compute absolute magnitudes in V, I, K, and Wesenheit W$_{VI}$ bandpasses corrected for interstellar extinction and Lutz-Kelker-Hanson bias. Adding our previous absolute magnitude determination for $\delta$ Cep, we construct  Period-Luminosity relations for ten Galactic Cepheids.

We compare our new Period-Luminosity relations with those adopted by several recent investigations, including the Freedman and Sandage H$_0$ projects. Adopting our Period-Luminosity relationship would tend to increase the Sandage H$_0$ value, but leave the Freedman H$_0$ unchanged.  Comparing our Galactic Cepheid PLR with those derived from LMC Cepheids, we find the slopes for K and W$_{VI}$ identical in the two galaxies within their respective errors.  Our data lead to a W$_{VI}$ distance modulus for the Large Magellanic Cloud, m-M = 18.50$\pm$0.03, uncorrected for any metallicity effects. Applying recently derived metalllcity corrections yields a corrected LMC distance modulus of (m-M)$_0$=18.40$\pm$0.05. Comparing our Period-Luminosity relationship to solar-metallicity Cepheids in NGC 4258 results in a distance modulus, 29.28 $\pm$ 0.08, which agrees with that derived from maser studies. 
\end{abstract}




%

\section{Introduction}

Many of the methods used to determine the distances to remote galaxies and ultimately the size, age, and shape of the Universe itself depend on our knowledge of the distances to local objects. Among the most important of these are the Cepheid variable stars. The Cepheid Period-Luminosity Relation (hereafter PLR) was first identified by Leavitt (Leavitt \& Pickering 1912). This has led to considerable effort to determine the absolute magnitudes, M$_V$, of these objects, as summarized in the comprehensive reviews by Madore \& Freedman (1992), Feast (1999), and Macri (2005). 

As summarized by Freedman \etal (2001) Cepheids are
among the brightest stellar distance indicators and a critical initial step on the `cosmic distance ladder'. These `standard candles' are relatively
young stars, found in abundance in spiral galaxies.
For extragalactic distance determinations many independent objects can be observed in a single galaxy, affording a reduction in distance modulus error. Their large amplitudes and characteristic (sawtooth)
light curve shapes facilitate their discovery and identification.
Lastly, the Cepheid PLR has a small scatter. In the I band, the dispersion amounts to only  $\sim$0.1 mag (Udalski et al. 1999). 
 
Given that the distances of all local Cepheids, except Polaris, are in excess of 250pc, most of the past absolute magnitude determinations have used indirect approaches, for example \citet{Gro00}, \cite{Lan99}, \citet{Fea97a}, \citet{Fea97}, \citet{Fea98a},  and \citet{Fea99}. Various authors, {\it e.g.} Gieren \etal (1993) used Cepheid surface brightness to estimate distances and absolute magnitudes. For Cepheid variables, these determinations are 
complicated by dependence of the absolute magnitudes on color index and possibly metallicity. Only recently have relatively high-precision trigonometric parallaxes  (${\sigma_{\pi}\over \pi} \le 10\%$) been available for a very few Cepheids (the prototype, $\delta$ Cep and Polaris) from {\it HIPPARCOS}  \citep{Per97}. More recently we have determined the parallax of $\delta$ Cep with Fine Guidance Sensor 3 (FGS 3) on {\it Hubble Space Telescope} (\HST) with ${\sigma_{\pi}\over \pi} \sim 5\%$ precision \citep{Ben02b}. 
Long-baseline ground-based interferometry has  recently provided radii and, through various  surface brightness methods, distances \citep{Nor02,Ker04,Ker06}.  

Our immediate goal is to determine trigonometric parallaxes  for an additional nine nearby fundamental mode Galactic Cepheid variable stars. Our  target selection consisted in choosing the nearest Cepheids (using \HIP~ parallaxes), covering as wide a period range as possible. These stars are in fact the brightest known Cepheids at their respective periods. Our new parallaxes provide distances and ultimately absolute magnitudes, $M$, in several bandpasses. Additionally, our investigation of the astrometric reference stars provides an independent estimation of the line of sight extinction to each of these stars, a contributor to the uncertainty in the absolute magnitudes of our prime targets.
These Cepheids, all with near solar metallicity, should be unafflicted by potential variations in absolute magnitude due to metallicity variations, {\it e.g.} \citet{Gro04,Mac06}. Adding our previously determined absolute magnitude for $\delta$ Cep \cite{Ben02b}, we establish V, I, K, and W$_{VI}$ Period-Luminosity Relationships (PLR) using ten Galactic Cepheids with average metallicity, $\langle$[Fe/H]$\rangle$=0.02, a calibration that can be directly applied to external galaxies whose Cepheids exhibit solar metallicity. 

We describe our astrometry using one of our targets, \lC, as an example throughout. This longest-period member of our sample provides marginal evidence for any possible PLR V-band non-linearity and, if included,  anchors our PLR slopes. Additionally it is the only one of our sample in the period range typically used to establish extragalactic distance moduli. Hence, its parallax value deserves as much external scrutiny as possible. We discuss (Section~\ref{OBS}) data acquisition and analysis; present the results of spectrophotometry of the astrometric reference stars required to correct our relative parallax to absolute (Section~\ref{SpecPhot}); derive absolute parallaxes for these ten Cepheid variable stars (Section 4); derive Cepheid absolute magnitudes (Section 5); and finally in Section 6 we determine a number of Period-Luminosity Relations, briefly discuss the possibility of
nonlinearity in the galactic V band PLR, discuss the distance scale ramifications of our results, and apply our PLR to two interesting cases - the LMC and NGC 4258. We summarize  in Section 7.

\section{Observations and Data Reduction}  \label{OBS}

\nocite{Nel03}Nelan \etal (2003) 
provides an overview of the
FGS instrument and \nocite{Ben02b} Benedict \etal (2002b) describe the fringe tracking (POS) mode astrometric capabilities 
of an FGS, along with the data acquisition and reduction strategies also used in the present study. 
We time-tag our data with a modified Julian Date, MJD  =  JD - 2400000.5, and abbreviate millisecond of arc, mas, throughout.

Eleven sets of astrometric data were acquired with {\it HST} FGS 1r for each of our nine new science targets. For details on our tenth, previously analyzed Cepheid, $\delta$ Cep, see \cite{Ben02b}. We obtained most of these eleven sets in pairs at maximum parallax factor typically separated by a week, a strategy designed to protect against unanticipated {\it HST} equipment problems. We encountered none, obtaining 110 orbits without the slightest difficulty. A few single data sets were acquired at various minimum parallax factors to aid in separating parallax and proper motion. Each complete data aggregate spans 1.49 to 1.95 years. Table~\ref{tbl-LOO} contains the epochs of observation, pulsational phase, and estimated B-V color index (required for the lateral color correction discussed in Section \ref{ASTMOD}) for each Cepheid. 
The B-V colors are inferred from color curves 
constructed from the Cepheid photometric database\footnote{
\url{http://ftp.sai.msu.su/groups/cluster/CEP/PHE}} cited 
by Berdnikov \etal (2000).
In the case of RT Aur, we supplemented the few data in 
that source with B-V values from Moffett \& Barnes 
(1984), Barnes \etal (1997), and Kiss (1998).  We 
adopted the periods and epochs listed by Szabados 
(1989, 1991) for these Cepheids except for \lC. Because the period of \lC~ changes unpredictably with time \citep{Sza89}, we derived a new period based upon the more recent V data 
in the Berdnikov \etal database.  The 
Cepheids $\eta$ Aql, $\zeta$ Gem, and X Sgr also have variable 
periods, but they vary quadratically and predictably. We took these variations into account when computing the 
phases.

\begin{figure}
\epsscale{0.75}
\plotone{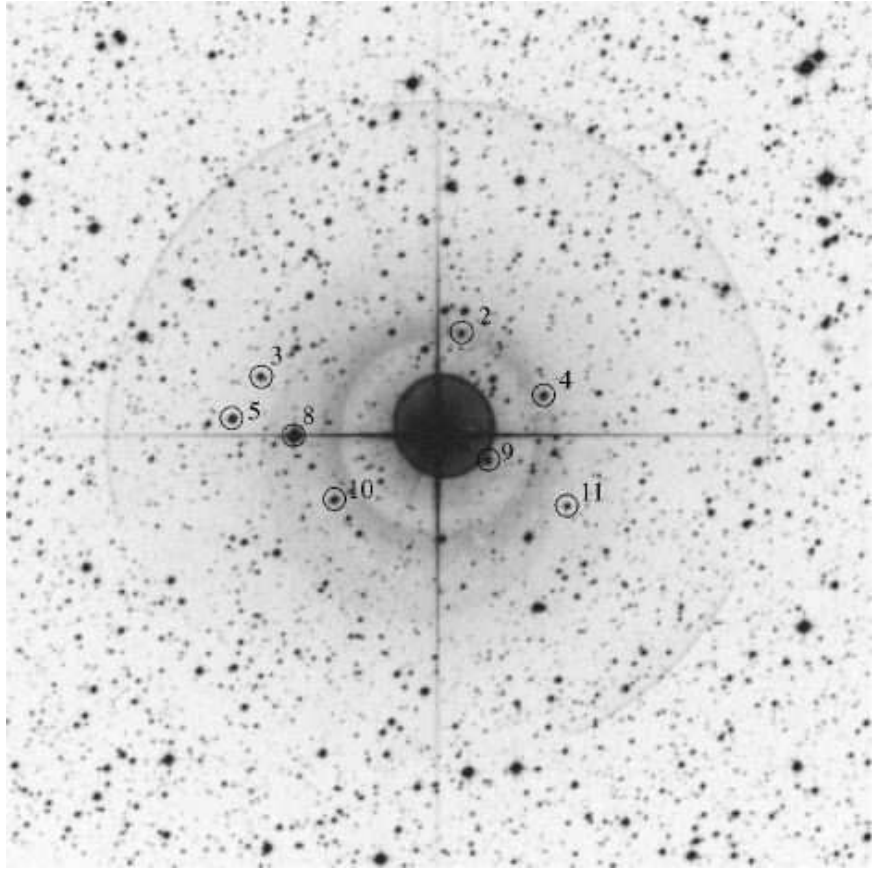}
\caption{\lC~ field with astrometric reference stars marked. Box is 15\arcmin ~across.}
\label{fig-1}
\end{figure}

Each individual data set required approximately 33 minutes of spacecraft time. The data were reduced and calibrated as detailed in  McArthur \etal (2001)\nocite{mca01}, Benedict \etal (2002a)\nocite{Ben02a}, Benedict \etal (2002b)\nocite{Ben02b}, and Soderblom \etal (2005). At each epoch we measured reference stars and the target multiple times, this to correct for intra-orbit drift of the type seen in the cross filter calibration data shown in figure 1 of Benedict \etal (2002a)\nocite{Ben02a}.  A typical distribution of reference stars  on a second generation Digital Sky Survey R image near one of our science targets (\lC) is shown in Figure \ref{fig-1}. The somewhat elongated distribution of reference stars is forced by the shape of the FGS field of view and the overlap area. The orientation of each successive observation at near-maximum parallax factor changes by 180\arcdeg, mandated by {\it HST} solar panel illumination constraints. 

Data are downloaded from the \HST~ archive and passed through a pipeline processing system. This pipeline extracts the astrometry measurements (typically one to two minutes of fringe position information acquired at a 40 Hz rate, which yields several thousand discrete measurements), extracts the median (which we have found to be the optimum estimator), corrects for the Optical Field Angle Distortion (McArthur \etal 2002), and attaches all required time tags and parallax factors.  

Table~\ref{tbl-AP} collects measured properties for our target Cepheids, including pulsational period, log of that period, $\langle$V$\rangle$ ,$\langle$I$\rangle$, $\langle$K$\rangle$, $\langle$B-V$\rangle$, E(B-V), A$_V$, and A$_K$. Photometry is from \cite{Gro99} and \cite{Ber96}. The $\langle$K$\rangle$ values for $\delta$ Cep 
and T Vul were corrected following Berdnikov (2006, private 
communication). Cepheid $\langle$K$\rangle$ is in the CIT system. The $\langle$I$\rangle$ is in the Cousins system. All reddening values are either derived from our reference star photometry or adopted from those listed in the David Dunlap Observatory Cepheid database \citep{Fer95}. Our reddening selection criterion is discussed in Section 5.

\setcounter{footnote}{0}
\section{Spectrophotometric Parallaxes of the Astrometric Reference Stars} \label{SpecPhot}
The following review of our astrometric and spectrophotometric techniques uses the \lC~ field as an example. Given that \lC~ has the longest period in our sample, it may have a significant effect on the slopes of the PLR we eventually construct. It also has a period most like that of extragalactic Cepheids used in distance determination.
Because the parallaxes determined for the Cepheids will be
measured with respect to reference frame stars which have their own
parallaxes, we must either apply a statistically derived correction from relative to absolute parallax (Van Altena, Lee \& Hofleit 1995, hereafter YPC95) or estimate the absolute parallaxes of the reference frame stars listed in Table \ref{tbl-POS}. In principle, the colors, spectral type, and luminosity class of a star can be used to estimate the absolute magnitude, M$_V$, and V-band absorption, A$_V$. The absolute parallax is then simply,
\beq
\pi_{abs} = 10^{-{(V-M_V+5-A_V)}\over5}
\eeq

The luminosity class is generally more difficult to estimate than the spectral type (temperature class). However, the derived absolute magnitudes are critically dependent on the luminosity class. As a consequence we use as much additional information as possible in an attempt to confirm the luminosity classes. Specifically, we obtain 2MASS\footnote{The Two Micron All Sky Survey
is a joint project of the University of Massachusetts and the Infrared Processing
and Analysis Center/California Institute of Technology } photometry and UCAC2 proper motions (Zacharias \etal 2004) for a one degree square field centered on each science target, and iteratively employ the technique of reduced proper motion (Yong \& Lambert 2003\nocite{Yon03}, Gould \& Morgan 2003\nocite{Gou03}, Ciardi 2004\nocite{Cia04}) to confirm our giant/dwarf classifications. 

\subsection{ Reference Star Photometry}
Our band passes for reference star photometry include: BVI (from recent measurements with the New Mexico State University 1m telescope for the northern Cepheids, and from the South African Astronomical Observatory (SAAO) 1m for the southern Cepheids) and JHK (from 2MASS).  For reference star spectrophotometric parallaxes only, the 2MASS JHK have been transformed to the Bessell \& Brett (1988) system using the transformations provided in Carpenter (2001)\nocite{Car01}. Table \ref{tbl-IR} lists BVIJHK photometry for targets and reference stars bright enough to have 2MASS measurements. In addition Washington-DDO photometry \citep{Pal94,Maj00} was used to confirm the luminosity classifications for the later spectral type reference stars.

\subsection{Reference Star Spectroscopy }
The spectra from which we estimated \lC~ reference star spectral type and luminosity class come from the South African Astronomical Observatory (SAAO) 1.9m telescope. Spectral classifications for the $\beta$ Dor and X, Y, and W Sgr fields were also provided by SAAO. The SAAO resolution  
was 3.5 \AA/ (FWHM) with wavelength coverage from 3750 \AA$\leq$ $\lambda$ 
$\leq$ 5500 \AA.  Spectroscopic classification of the reference stars in the fields of
RT Aur and $\zeta$ Gem was accomplished using data obtained with the
Double Imaging Spectrograph (DIS) on the Apache Point Observatory 3.5 m 
telescope\footnote{The
Apache Point Observatory 3.5 m telescope is owned and operated by
the Astrophysical Research Consortium.}. We used the high resolution gratings, delivering a dispersion 
of 0.62 \AA /pix, and covering the wavelength range of 3864 $\leq$ $\lambda$ 
$\leq$ 5158 \AA. Spectroscopy of the reference stars in the fields of Y Sgr, 
FF Aql, $\eta$ Aql, and T Vul was obtained using the R-C Spectrograph on
the KPNO 4 m. The ``t2kb'' detector with grating ``\#47'' was used to
deliver a dispersion of 0.72 \AA /pix, covering the wavelength range
3633 $\leq$ $\lambda$ $\leq$ 5713 \AA. 
Classifications used a combination of template matching and line ratios. Spectral types for the stars are generally better than $\pm$2
subclasses. 

\subsection{Interstellar Extinction} \label{AV}
\begin{figure}
\epsscale{1.00}
\plotone{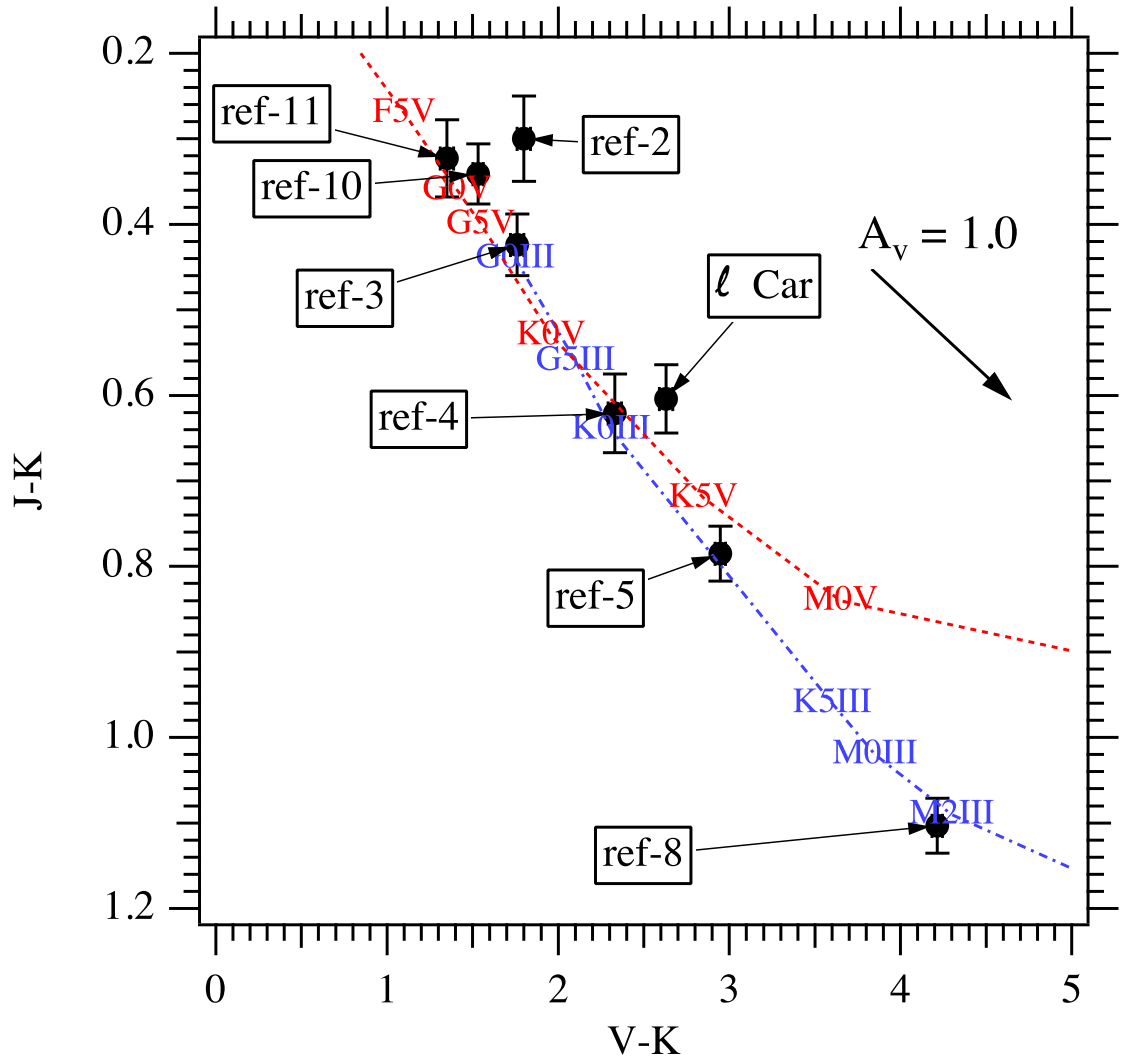}
\caption{J-K vs V-K color-color diagram for \lC~ and reference stars. The dashed line is the locus of  dwarf (luminosity class V) stars of various spectral types; the dot-dashed line is for giants (luminosity class III). The reddening vector indicates A$_V$=1.0 for the plotted color systems. For this low-Galactic latitude field $\langle A_V\rangle$ = 0.52 $\pm$ 0.06 magnitude (Table~\ref{tbl-AV}).}
\label{fig-2}
\end{figure}

To determine interstellar extinction we first plot the reference stars on a J-K vs. V-K color-color diagram. A comparison of the relationships between spectral type and intrinsic color against those we measured provides an estimate of reddening. Figure \ref{fig-2} contains the \lC~ J-K vs V-K color-color diagram and reddening vector for A$_V$ = 1.0. Also plotted are mappings between spectral type and luminosity class V and III from Bessell \& Brett~(1988)\nocite{Bes88} and Cox (2000)\nocite{Cox00}. Figure~\ref{fig-2}, along with the estimated spectral types, provides an indication of the reddening for each reference star. 

Assuming an R = 3.1 Galactic reddening law (Savage \& Mathis 1979\nocite{Sav79}), we derive A$_V$ values by comparing the measured colors (Table \ref{tbl-IR} ) with intrinsic (V-K)$_0$, (B-V)$_0$, (U-B)$_0$, (J-K)$_0$, and (V-I)$_0$, colors from Bessell \& Brett~(1988)\nocite{Bes88} and Cox (2000). We estimate A$_V$ from A$_V$ = 1.1E(V-K) = 5.8E(J-K) = 2.77E(U-B) = 3.1E(B-V) = 2.26E(V-I), where the ratios of total to selective extinction were derived from the Savage \& Mathis (1979) reddening law and a reddening estimate in the direction of \lC~ from \cite{Sch98}, via NED\footnote{NASA/IPAC Extragalactic Database}.  All resulting A$_V$ are collected in Table~\ref{tbl-AV}. We then calculate a field wide average A$_V$ to be used in equation 1. For the \lC~ field $\langle A$$_V$$\rangle$ = 0.52$\pm$0.06 magnitude. In this case our independent determination is in good agreement with the David Dunlap Observatory online Galactic Cepheid database\footnote{http://www.astro.utoronto.ca/DDO/research/cepheids/cepheids.html}, which averages seven measurements of color excess to obtain $\langle$E(B-V)$\rangle$ = 0.163$\pm$0.017 , or $\langle A$$_V$$\rangle$ = 0.51$\pm$0.05.
 
Using the \lC~ field as an example, we find that the technique of reduced proper motions can provide a possible confirmation of reference star estimated luminosity classes. The precision of existing proper motions for all the reference stars was $\sim$5 mas y$^{-1}$, only suggesting discrimination between giants and dwarfs. Typical errors on H$_K$, a parameter equivalent to absolute magnitude, M$_V$, were about a magnitude. Nonetheless, a reduced proper motion diagram did suggest that ref-4, -5, and -8 are not dwarf stars. They are considerably redder in J-K than the other stars in the present program classified as dwarfs. Giants are typically redder in J-K than dwarfs for a given spectral type (Cox 2000). Our luminosity class uncertainty is reflected in the input spectrophotometric parallax errors (Table~\ref{tbl-SPP}). We will revisit this additional test in Section~\ref{ASTMOD}, once we have higher precision proper motions obtained from our modeling.

\subsection{Estimated Reference Frame Absolute Parallaxes}

We derive absolute parallaxes for each reference star using M$_V$ values from \cite{Cox00} and the $\langle$A$_V\rangle$ derived from the photometry. Our adopted errors for (m-M)$_0$ are 0.5 mag for all reference stars. This error includes uncertainties in $\langle A$$_V$$\rangle$ and the spectral types used to estimate M$_V$.
Our reference star parallax estimations from Equation 1 are listed in Table \ref{tbl-SPP}.
For the \lC~ field individually, no reference star absolute parallax is better determined than ${\sigma_{\pi}\over \pi}$ = 23\%. The average absolute parallax for the reference frame is $\langle\pi_{abs}\rangle = 0.85$ mas.
We compare this to the correction to absolute parallax discussed and presented
in YPC95 (section 3.2, fig. 2). Entering
YPC95, fig. 2, with the \lC~ Galactic
latitude, l = -7\arcdeg, and average magnitude for the
reference frame, $\langle$V$_{ref}$$\rangle$= 13.0, we obtain a correction
to absolute of 1 mas. This gives us confidence in our spectrophotometric determination of the correction to absolute parallax. As in past investigations we prefer to introduce into our reduction model our spectrophotmetrically estimated reference star parallaxes as observations with error. The
use of spectrophotometric parallaxes offers a more direct (less Galaxy model-dependent) way of
determining the reference star absolute parallaxes.

\section{Absolute Parallaxes of Galactic Cepheids}

\subsection{The Astrometric Model} \label{ASTMOD}

With the positions measured by FGS 1r we determine the scale, rotation, and offset ``plate
constants" relative to an arbitrarily adopted constraint epoch (the so-called ``master plate") for
each observation set (the multiple observtions of reference stars and Cepheid targets acquired at each epoch listed in Table 1). The rotation to the sky of the master plate is initially set at a value provided by the \HST~ ground system. The mJD of each observation set is listed in Table~\ref{tbl-LOO}, along with a Cepheid B-V estimated from a phased light curve. Our \lC~ reference frame contains 6 stars. All the Cepheid primary science targets, including \lC, are bright enough to require the use of the FGS neutral density filter.  Hence, we use the modeling approach outlined in Benedict \etal (2002b)\nocite{Ben02b}, with corrections for both cross-filter and lateral color positional shifts, using values specific to FGS 1r determined from previous calibration observations with that FGS. 

We employ GaussFit (Jefferys \etal 1988)\nocite{Jef88} to minimize $\chi^2$, our model goodness-of-fit metric. GaussFit has a number of features, including a complete programming language
designed especially to formulate estimation problems, a built-in compiler and interpreter to support
the programming language, and a built-in algebraic manipulator for calculating the required partial
derivatives analytically. The program and sample models are freely available\footnote{http://clyde.as.utexas.edu/Software.html}. 

The solved equations
of condition for the \lC~ field are:
\beq
        x'  =  x + lc_x(\it B-V) {- \Delta XFx}
\eeq
\beq
        y'  =  y + lc_y(\it B-V) { - \Delta XFy}
\eeq
\beq
\xi = Ax' + By' + C  - \mu_x \Delta t  - P_\alpha\pi_x
\eeq
\beq
\eta = Dx' + Ey' + F  - \mu_y \Delta t  - P_\delta\pi_y
\eeq

where $\it x$ and $\it y$ are the measured coordinates from {\it HST};
$\it lc_x$ and $\it lc_y$ are the
lateral color corrections; $\Delta$XFx and $\Delta$XFy are the cross filter corrections in $\it x$ and $\it y$, applied only to the observations of each Cepheid; and $\it B-V $ are
the  B-V  colors of each star. A,  B, D, and E   
are scale and rotation plate constants, C and F are
offsets; 
$\mu_x$ and $\mu_y$ are proper motions; $\Delta$t is the epoch difference from the mean epoch;
$P_\alpha$ and $P_\delta$ are parallax factors;  and $\it \pi_x$ and $\it \pi_y$
 are  the parallaxes in x and y. $ x'$ and $y'$ are FGS positions corrected for lateral color and cross-filter shifts. $\xi$ and $\eta$ are relative positions in arcseconds.   We obtain the parallax factors from a JPL Earth orbit predictor (Standish 1990\nocite{Sta90}), upgraded to version DE405.

There are additional equations of condition relating an initial value (an observation with associated error) and final parameter value. There is one such equation in the model for each parameter of interest: reference star and target color index,  proper motion, and (excepting the Cepheid target) spectrophotometric parallax.   Through these additional equations of condition the $\chi^2$ minimization process is allowed to adjust parameter values by amounts constrained by the input errors. We also similarly adjust the lateral color parameters, master plate roll, and cross filter parameters. The end results are the final values of the  parameters of intertest. In this quasi-Bayesian approach prior knowledge is input as an observation with associated error, not as a hardwired quantity known to infinite precision.  

\begin{figure}
\epsscale{0.75}
\plotone{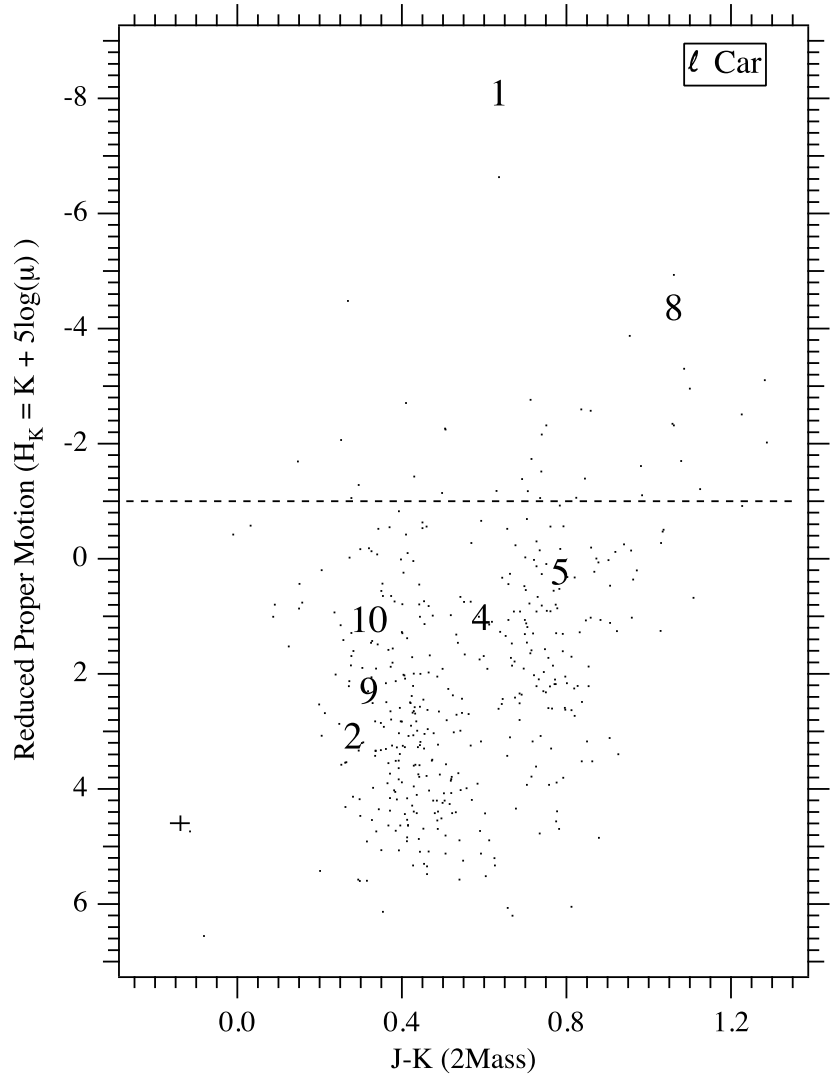}
\caption{Reduced proper motion diagram for 430 stars in a $1/3$ degree field centered on  \lC~. Star identifications are shown for \lC~ (`1') and our astrometric reference stars. H$_K$ for these stars is calculated using our final proper motions (Table~\ref{tbl-PM}). For a given spectral type giants and sub-giants have more negative H$_K$ values and are redder than dwarfs in J-K. Reference stars ref-4, -5, and -8 are confirmed to be giant stars. The cross in the lower left corner indicates representative errors along each axis.} \label{fig-3}
\end{figure}

For example input proper motion values  have typical errors of 4--6 mas y$^{-1}$ for each coordinate.    Final proper motion values and errors obtained from our modeling of {\it HST} data for the \lC~ field are listed in Table~\ref{tbl-PM}.  Adjustments to the proper motion estimates required to minimize $\chi^2$ averaged 3 mas yr$^{-1}$. For completeness, transverse velocities, given our final parallaxes, are listed in Table~\ref{tbl-PiVt}. As a final test of the quality of our prior knowledge of reference star luminosity class listed in Table~\ref{tbl-SPP}, we employ the technique of reduced proper motions. We obtain proper motion and J, K photometry from UCAC2 and 2MASS for a ${1\over3}$\arcdeg $\times$ ${1\over3}$\arcdeg~ field centered on \lC. Figure~\ref{fig-3} shows H$_K$ = K + 5log($\mu$) plotted against J-K color index for 436 stars. If all stars had the same transverse velocities, Figure~\ref{fig-3} would be equivalent to an HR diagram. \lC~ and reference stars are plotted as ID numbers from Table~\ref{tbl-PM}. \lC~ is `1' in Figure~\ref{fig-3}. With our precise proper motions (Table~\ref{tbl-PM}) errors in H$_K$ are now $\sim0.3$ magnitude. Reference stars ref-4, -5, and -8 remain clearly separated from the others, supporting their classification as giants. 

We stress that for no Cepheid in our program was a previously measured parallax used as prior knowledge and entered as an observation with error. Only reference star prior knowledge was so employed. Our Cepheid parallax results are blind to previous parallax measures from \HIP~ and/or parallaxes from surface brightness estimates.

\subsection{Assessing Reference Frame Residuals}
\begin{figure}
\epsscale{0.5}
\plotone{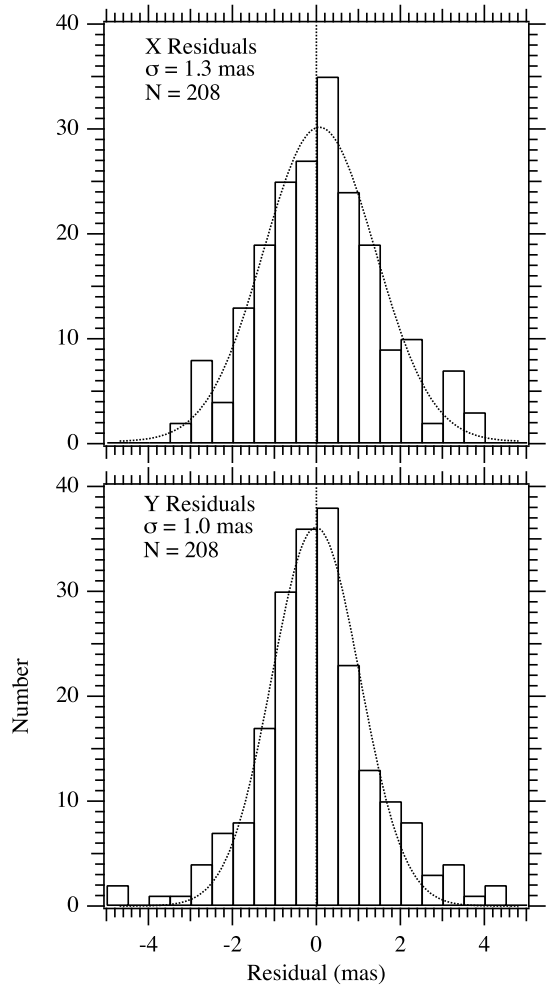}
\caption{Histograms of x and y residuals obtained from modeling \lC~ and astrometric reference stars with equations 4 and 5, constraining D=-B and E=A. Distributions are fit with gaussians whose 1-$\sigma$ dispersions are noted in the plots.} \label{fig-4}
\end{figure}

The Optical Field Angle Distortion calibration (McArthur \etal 2002\nocite{McA02}) reduces as-built {\it HST} telescope and FGS 1r distortions with amplitude $\sim1\arcsec$ to below 2 mas over much of the FGS 1r field of regard.  From histograms of the \lC~ field astrometric residuals (Figure~\ref{fig-4}) we conclude that we have obtained satisfactory correction. The resulting reference frame `catalog' in $\xi$ and $\eta$ standard coordinates (Table \ref{tbl-POS}) was determined
with	average position errors $\langle\sigma_\xi\rangle= 0.50$	 and	$\langle\sigma_\eta\rangle = 0.62$ mas.

To determine if there might be unmodeled - but possibly correctable -  systematic effects at the 1 mas level, we plotted  reference frame X and Y residuals against a number of spacecraft, instrumental, and astronomical parameters. These included X, Y position within our total field of view; radial distance from the field of view center; reference star V magnitude and B-V color; and epoch of observation.  We saw no obvious trends. 

\subsection{Absolute Parallaxes of the Cepheids} \label{AbsPi}
For the \lC~ and Y Sgr fields  we reduced the number of modeling coefficients in equations 3 and 4 to four, as done for our previous work on the Pleiades \citep{Sod05}. We constrained the fit to have a single scale term by imposing D = -B and E = A. Final model selection for all fields was based on reference star placement relative to the target, total number of reference stars, reduced $\chi^2$ ($\chi^2$/DOF, where DOF = degrees of freedom), and parallax error.  Absolute parallaxes, relative proper motions, and transverse velocities for \lC~ and associated reference stars are collected in Tables~\ref{tbl-PM} and \ref{tbl-PiVt}. Parallaxes for all Cepheids are collected in Table~\ref{tbl-SUM}.

All our Cepheid parallaxes directly rely on the estimates of reference star parallaxes. Should anyone wish to independently verify our results, the reference stars used in this study are all identified in archival material\footnote{\url http://www.stsci.edu/observing/phase2-public/9879.pro} held at the Space Telescope Science Institute. Adopted reference star spectral types and the parallaxes resulting from our modeling are listed in Table~\ref{tbl-allP}. Similar data for the $\delta$ Cep reference stars can be found in \cite{Ben02b}.

\subsection{{\it HST} Parallax Accuracy}
Our parallax precision, an indication of our internal, random error, is $\sim$ 0.2 mas. To assess our accuracy, or external error, we  have compared (Benedict \etal 2002b,  Soderblom \etal 2005) our parallaxes with results from independent measurements from HIPPARCOS (Perryman \etal 1997).  Other than for the Pleiades (Soderblom \etal 2005), we  have no large systematic differences with {\it HIPPARCOS} for any objects with ${\sigma_{\pi}\over\pi}<$10\%. The next significant improvement in geometrical parallaxes for Cepheids will come from the space-based, all-sky astrometry missions {\it GAIA} \citep{Mig05} and {\it SIM} \citep{Unw05} with $\sim10~ \mu$arcsec precision parallaxes. Final results are expected by the end of the next decade.

\subsection{The Binary Cepheids} \label{BinCep}
Many of our target Cepheids have companions discovered spectroscopically with IUE ({\it c. f. } Evans 1995\nocite{Eva95}). Two of these, W Sgr \citep{Pet04} and FF Aql \citep{Eva90}, have published spectroscopic orbital elements. For these two targets we introduced the known orbital elements as observations with error and solve for inclination and perturbation size
as outlined in \cite{Ben02b}, using equations 6 and 7 from that paper.

Our results for W Sgr and FF Aql are summarized in Table~\ref{tbl-10}.
With the perturbation orbit semimajor axis, $\alpha$, the measured inclination, $i$, and an estimate of the secondary mass, we can estimate the mass of each Cepheid.  The secondary mass is estimated from the spectral type and a recent Mass-Luminosity relationship \citep{Hen04}. An improvement in the W Sgr mass is expected shortly, once the secondary spectral type is more tightly constrained (Evans \& Massa 2007). The major contributor to the FF Aql mass error is the parallax error.

One of our original targets, $\eta$ Aql, is thought to
be a binary from IUE spectra (B9.8 companion, Evans 1991).
As shown in Table~\ref{tbl-10}, we have successfully included perturbation
orbits for two other Cepheids, FF Aql and W Sgr,
simultaneously solving for parallax, proper motion, inclination,
and perturbation orbit semimajor axis.
However, spectroscopic orbital parameters
are fairly well known for those stars,
definitely not the case for $\eta$ Aql. With no period,
eccentricity, or periastron timing constraints from
previous radial velocity observations, and effectively only
five distinct epochs of astrometry, we cannot
determine a perturbation orbit. Our astrometry is clearly affected
by an, as yet unmodelable, motion.
Therefore,   $\eta$ Aql  cannot be included in this analysis. Ultimately, additional HST observations may serve to characterize
the (likely face-on) perturbation orbit, resulting in a usable parallax.

\section{The Absolute Magnitudes of the Cepheids} 
When using a trigonometric parallax to estimate the absolute
magnitude of a star, a correction should be made for the
Lutz-Kelker  bias \citep{Lut73} as modified by Hanson (1979). We justify the application of Lutz-Kelker-Hanson (LKH) with an appeal to Bayes Theorem. See \cite{Bar03}, section 4, for an accessible introduction to Bayes Theorem as applied to astronomy. Invoking Bayes Theorem to assist with generating absolute magnitudes from our  Cepheid parallaxes, one would say, "what is the probability that a star from this
population with this position would have parallax $\pi$ (as a function of $\pi$),
given that we haven't  yet measured $\pi$?"  In practice one would use the
space distribution of the population to which the star presumably belongs.
This space distribution is built into the prior p($\pi$) for $\pi$, and used to determine
\beq
 p(\pi | \pi_{observed} \& K) \sim p(\pi_{observed} | \pi \& K) p(\pi | K)
\eeq
where K is prior knowledge  about the space distribution of the
class of stars in question and `\&' is an `and' operator. The function p($\pi_{observed} | \pi$ \& K) is the
standard likelihood function, usually a gaussian normal with variance $\sigma_{\pi}$
The
"standard" L-K correction has p($\pi |$ K) $\sim\pi^{-4}$. Looking at a
star in a disk population close to the galactic plane requires $\pi^{-3}$
(ignoring spiral structure), which is the prior we use.
The LKH bias is proportional to $(\sigma_{\pi}/\pi)^2$. Presuming that all Cepheids in Table 2 belong to the same class of object  (evolved Main Sequence stars), we scale the LKH correction determined in Benedict \etal (2002b) for $\delta$ Cep and obtain the LKH bias corrections listed in Table~\ref{tbl-SUM}. For \lC~ we find LKH = -0.08 magnitude. The average LKH bias correction for all Cepheids in this study was -0.06 magnitude. We identify the choice of prior for this bias correction as a possible contributor to systematic errors in the zero points of our PLR, at the 0.01 magnitude level.

With $\langle$V$\rangle$= 3.724  (Table~\ref{tbl-AP}) and  given the  absolute parallax, \lCpi mas from Section \ref{AbsPi}, we determine a distance modulus for \lC. From Table \ref{tbl-AV} (Section \ref{AV}) we obtain a derived field-average absorption, $\langle$A$_V$$\rangle$ = 0.52. With this $\langle$A$_V\rangle$, the measured distance to \lC, and the LKH correction we obtain M$_V = -5.35\pm 0.22$ and a corrected true distance modulus,  (m-M)$_0$ = 8.56.  The M$_V$ error has increased slightly by combining the $\langle$A$_V$$\rangle$ error and the raw distance modulus error in quadrature. The M$_K$ values in Table~\ref{tbl-SUM} have slightly lower errors because the $\langle$A$_K$$\rangle$ values are lower with correspondingly lower errors to add in quadrature. 
The Wesenheit magnitude, W$_{VI}$, listed in Table~\ref{tbl-SUM} is the prescription of \citet{Freed01}, $W_{VI} = V - {A_V\over{E(V-I)}}(V-I)$. For the reddening law adopted by them $W_{VI} = V - 2.45(V-I)$. 

Results, including all proper motions, and absorption- and LKH bias-corrected absolute magnitudes, for the Cepheids in our program (except $\eta$ Aql) are collected in Table~\ref{tbl-SUM}. In half the cases the reddening values we derived from our reference star photometry agreed with that listed in the David Dunlap Observatory Cepheid database. For $\delta$ Cep and X Sgr we adopted a reddening derived as described in Section 3.3 because the photometry showed very consistent star-to-star reddening. For $\beta$ Dor, Y Sgr, and FF Aql we adopted the DDO color excess and an absorption A$_V$=3.1E(B-V) because the star-to-star reddening indicated extremely patchy absorption. Adopted absorption in the V and K band are listed in Table~\ref{tbl-AP}.

\section{Period-Luminosity Relations, Distance Scale Implications, and Applications}
\subsection{Period-Luminosity Relations from \HST ~Parallaxes}
Plotting the absorption and LKH bias corrected V, I, K and Wesenheit absolute magnitudes, M$_V$, M$_I$, M$_K$, and M$_{W(VI)}$, from Table~\ref{tbl-SUM} against the logarithm of the period ($logP$, Table~\ref{tbl-AP}) we obtain the Period-Luminosity relationships (PLR) contained in Figure~\ref{fig-5}. We parameterize all PLR as $M_X = a + b(logP - 1)$. Hence, the zero-points are for a Cepheid with logP = 1.   Our intercepts and slopes (B07) with $1\sigma$ errors  are collected in Table~\ref{tbl-11}, along with other recent determinations; Freedman \etal (2001, F01), Sandage \etal (2004, S04),  and Barnes \etal (2003, B03). Note that the Sandage M$_{W(VI)}$ was derived from their V and I PLR (\citet{San04}, eqs. 17 and 18), using  $W_{VI} = V -2.52(V-I)$. Adopting the Freedman reddening coefficient (2.45) would change the slope of the Sandage $W_{VI}$ PLR only by $+0.02$ magnitude.
 
The standard deviation of our residuals (Figure~\ref{fig-5}) are 0.10 magnitude for  M$_V$ and M$_I$, and 0.09 magnitude for M$_K$ and M$_{W(VI)}$. In each case the largest residual is that of W Sgr. Note that the determination of the W Sgr parallax  was complicated by the inclusion of a binary perturbation orbit. However, excluding W Sgr from the fit of, for example, M$_K$, changes the slope and intercept by less than 0.01 magnitude. 

Given the diversity of opinion regarding the applicability of LKH bias corrections (e.g. Smith 2003), even among the present author list, one of us (MF) suggested the following. The absolute magnitude error depends on the fractional parallax error, $\sigma_\pi/\pi$. When forming the PLR in Figure~\ref{fig-5}, a star, which has by chance an overestimated
parallax, will have greater weight in the solution than the same
star with an underestimated parallax. Feast (1998b), table 1, presents an extreme example of this. A correction based on  $\sigma_\pi/\pi$ seems required. We first fit a PLR with absolute mags {\em uncorrected for  LKH bias} and weight the stars by $({\pi \over \sigma})^2$. From the deviations of each star
from this PLR we deduce the parallax ($\pi_1$) which each star would have to have
to fall on the PLR. We then redo the PLR, weighting the uncorrected absolute mags
by $({\pi_1 \over \sigma})^2$. Changes in the PLR slope and intercept are less than 0.005 within three iterations.  Table~\ref{tbl-11}  lists PLR slopes and intercepts obtained with this particular weighting scheme as B07f. We note that the B07f slopes and intercepts agree within their respective errors with those obtained employing LKH bias corrections (B07).

\subsection{Pulsation Modes and Shocks}
There is always a possibility that a
Cepheid will be pulsating in an overtone, especially at shorter periods. The ratio
of the fundamental period $P_{0}$ to the overtone period $P_{1}$
is given by :
\begin{equation}
 P_{1}/P_{0} = 0.720 - 0.027\log P_{0}
\end{equation}
({\it e.g.} Alcock \etal  1995).
Thus for the $M_{V}$ - $log P$  plot found in Figure~\ref{fig-5} 
 overtones will lie about
0.4 magnitude above fundamental pulsators at a given period. As Figure~\ref{fig-5} shows
there is no evidence that any of the Cepheids in our sample deviate
by this amount from a single relation. This is particularly interesting
in the case of FF Aql. This low amplitude ($\Delta V = 0.33$) Cepheid
has been classed as an overtone pulsator on the basis of fourier
analyses of both the light curve (Antonello \etal 1990) and the velocity curve
(Kienzle \etal 1999). It was
nevertheless classed as a fundamental pulsator by
 Sachkov (1997) from a Baade-Wesselink type
radius estimate. None of the other Cepheids in our sample
have been suggested to be overtone pulsators so far as we are
aware, although $\zeta$ Gem has a relatively low amplitude
($\Delta V = 0.49$).
As shown in Table~\ref{tbl-SUM} our absolute magnitude
for FF Aql ($\log P = 0.6504$) is $M_{V} = -3.05 \pm 0.15$.
This is $0.15$ magnitude fainter than the mean relation (Figure~\ref{fig-5}).
If it were an overtone pulsator it would be $0.53$ magnitude fainter
than expected. 
We therefore conclude that it is not an 
overtone pulsator. This results suggests that despite the rather
clear division of light and velocity fourier coefficients of
Cepheids
into two groupings, this may not always correspond to a division
into fundamental and overtone pulsators. 

Mathias et al. (2006) find that X Sgr is apparently unusual
in showing evidence of multiple shock waves in its atmosphere.
It is therefore worth noting that this Cepheid appears
quite normal as far as its position in any of our PLR is concerned.

\subsection{Distance Scale Implications}
In this section we use two methods to compare our new parallaxes and our derived PLR with those from previous investigations. The first approach is simply to compare zero-points and slopes of various PLR. These will be flagged as {\bf DC} (direct comparison). The second approach uses reduced parallaxes \citep{Fea02} to solve for independent zero-points, which are then compared with our new values listed in Table~\ref{tbl-11}. This approach is denoted {\bf RP} (reduced parallaxes). In the following we carry out these comparisons with two populations of Cepheids, first Galactic Cepheids, then Cepheids in the Large Magellanic Cloud (LMC).

\subsubsection{Comparison with Other Galactic PLR}
6.3.1.1 - Barnes \etal (2003): develop and describe a Bayesian statistical analysis to solve the surface brightness equations for Cepheid distances and stellar properties. Their analysis
averages over the probabilities associated with several models rather than attempting to pick the ÔÔ best model ÕÕ from several possible models. They obtain a PLR using a sample of 13 Galactic Cepheids. 

 Before comparing their V band PLR with ours, we 
can compare our parallax for T 
Vul with that determined by Barnes et al. (2003) 
using a Bayesian solution in the visual surface 
brightness technique.  As discussed by Barnes \etal the 
surface brightness technique determines a quasi-geometric 
parallax rather than a distance.  Their table 7 gives a 
parallax for T Vul of $1.65 \pm 0.11$ mas, compared 
to our \HST~ parallax of $1.90 \pm 0.23$ mas.  

However, the surface brightness parallax contains an unknown systematic uncertainty, because it depends upon the 
adopted factor for conversion from radial velocity to 
pulsational velocity, normally denoted $p$.  (The larger 
the value of $p$, the smaller the parallax computed.)  The 
value adopted by Barnes \etal for T Vul is $p=1.37$.  Recent values in the literature, appropriate for periods near 
that of T Vul, range from $p=1.27 \pm 0.06$ (Merand et al. 2005) 
to $p=1.48 \pm 0.21$ (Gieren et al. 2005).  We can use the HST 
parallax to infer a quasi-geometrical value of $p$ by demanding that 
the surface brightness parallax match the \HST~ parallax. The result 
is $p=1.19 \pm 0.16$.  This value is consistent with the geometrically 
determined value of Merand et al., with the canonical value $p=1.36$ 
that is often used for Cepheids, and with the value $p=1.48$ from 
Gieren et al. (2005) given the mutual uncertainties.  Even though the 
uncertainty is larger than we would like, our value is only the 
second geometrically or quasi-geometrically determined value of 
$p$ after that of Merand et al. (2005).

{\bf DC} A comparison of the Barnes \etal PLR  in 
$V$ (based on thirteen Galactic Cepheids) with that for our ten-
Cepheid solution is quite satisfactory.  As shown in 
Table ~\ref{tbl-11} the zero points and the slopes both agree within 
$1\sigma$.  The agreement would have been slightly 
better had the reddening law chosen by Barnes  \etal been 
the same as adopted in this work.  Their choice of $R$ $\sim$3.35
leads to larger values of $A_v$ than does the law adopted 
here.  Adjusting the surface brightness $M_V$ values to 
our reddening law would change the slope determined by 
Barnes  \etal about $0.5\sigma$ closer to the slope 
determined in the present work. 

6.3.1.2 -  Freedman and Sandage:  
An ultimate goal of many workers has been to use Cepheids to
establish the Hubble Constant, $H_{0}$. Two major groups are recently
or presently involved in this effort and their results are summarized in
Freedman \etal (2001) and Sandage \etal (2006).  Here we investigate
how our new parallaxes test some of their basic assumptions. 

Both groups effectively use `reddening free'
(W$_{VI}$ type) relations. However they differ, among other things,
by use of different reddening laws and therefore different color
coefficients in their relations. The relation used by Freedman
\etal (2001) is equation 5 of their section 3.3 and is derived
from the OGLE LMC work on Cepheids (Udalski \etal 1999).
With the other equations in that section it can be written
\begin{equation}
Mod = V + 3.255(log P-1) -2.45(V-I) + F
\end{equation}
where the Freedman zero-point, $F =+ 5.899 +A$, is based on
an LMC distance modulus of 18.50 and a correction, A, for any 
metallicity difference with the
LMC. The relation between A and metallicity adopted by Freedman
\etal leads to $A = +0.08$ in the case of metal normal
(Galactic) Cepheids. 

Sandage \etal adopt different relations
for the LMC and our Galaxy. For our Galaxy they adopt
\begin{equation}
Mod = V + 3.746(logP -1) - 2.523(V-I) + S
\end{equation}
where the Sandage zero-point $S = +5.959$ 
is based on data from Baade-Wesselink type analyses and from Cepheids in
Galactic clusters. 

{\bf RP} Our first test  uses the method of reduced parallaxes
outlined in Feast (2002, eqs 1, 2, and 3) to estimate the zero-points, F and S,
in equations 8 and 9. The results are shown in Table~\ref{tbl-RPar} together
with the values adopted by the two groups. In our reductions we have 
assumed that all the uncertainty is in the parallaxes. The small 
scatter about the PLR relations discussed in this paper suggest
that other sources of uncertainty are small. This assumes 
(equation 3 of Feast 2002) that the uncertainty in the
magnitude and the intrinsic scatter in the adopted relation
are small. If they were significant they would
decrease the value of the derived constants
in the above Equations 9 and 10 (decrease the absolute
brightness of the Cepheids) due to a change in the relative weights
of the stars. For instance in the case of S this changes from
+5.96 to +5.93 if each of the above uncertainties was 0.07 mag and to
+5.91 if they were both 0.1 mag. The results in  Table~\ref{tbl-RPar} show that the 
reduced parallax calculation gives a zero-point for the Sandage 
Galactic relation in agreement with that adopted by them, whereas 
the Freedman zero-point is 0.16 mag smaller than the one used by 
them.

{\bf DC} For our second test  we calculate parallaxes for our 10 Cepheids using Equations 8 and 9 with the Freedman and Sandage zero-points. These `post-diction' parallaxes are listed in Table~\ref{tbl-parPost}.  
The results in Table~\ref{tbl-parPost} confirm the differences found in the {\bf RP} test. The small 
differences between Tables \ref{tbl-RPar} and \ref{tbl-parPost} are due to the fact that the {\bf DC} test (Table~\ref{tbl-parPost}) was carried out using unweighted quantities

 In summary, using {\bf RP} and our Table~\ref{tbl-SUM} parallaxes to derive the zero-points of Equations
8 and 9, we find an agreement with Sandage et al. (Eq 9) within
the errors but a difference of $0.16 \pm 0.04$ from Freedman et al.
(Eq 8).
However most HST-based work
on extragalactic Cepheids has been heavily weighted to the
longer periods, true both of the Freedman \etal
and  Sandage \etal programs. Thus it is important to note (Table~\ref{tbl-parPost})
that for our longest period Cepheid, $l$ Car, the Freedman \etal
relation predicts a parallax in better agreement with ours
than does that of  Sandage \etal. This {\bf DC} result is shown in
a slightly different way in Fig~\ref{fig-6} where we plot both the
Freedman et al. and the Sandage \etal Wesenheit M$_{W(VI)}$ PLR together with
those derived from our data. We allow both the coefficient of
(logP - 1) and the zero-point to vary, but adopted the relevant
color coefficient. These plots show that at the longer periods (log P $>$ 1) relevant to much extragalactic work the Freedman  \etal relation lies close to our best estimate and implies little change to
their derived H$_0$, despite its being inconsistent with our
data at shorter periods. On the other hand
using our our M$_{W(VI)}$ PLR (Fig~\ref{fig-6})
would tend to increase the Sandage estimate of H$_0$,
at least where it depends on galaxies of near solar metallicity. This
result depends crucially on \lC. It would clearly be important to strengthen the long period calibration by obtaining additional high-precision parallaxes of long-period Cepheids.

\subsubsection{Comparison with LMC Cepheid PLR} \label{LMCcomp}
 To carry out direct comparisons of Galactic and LMC PLR ({\bf DC}) it is necessary to compare PLR with equal slopes, because the logP=1 intercept depends on PLR slope.  Where necessary we have refit our PLR, constraining the slope to those established for LMC Cepheids, to re-determine zero-points, all uncorrected for metallicity effects.
 
\begin{figure}
\epsscale{0.85}
\plotone{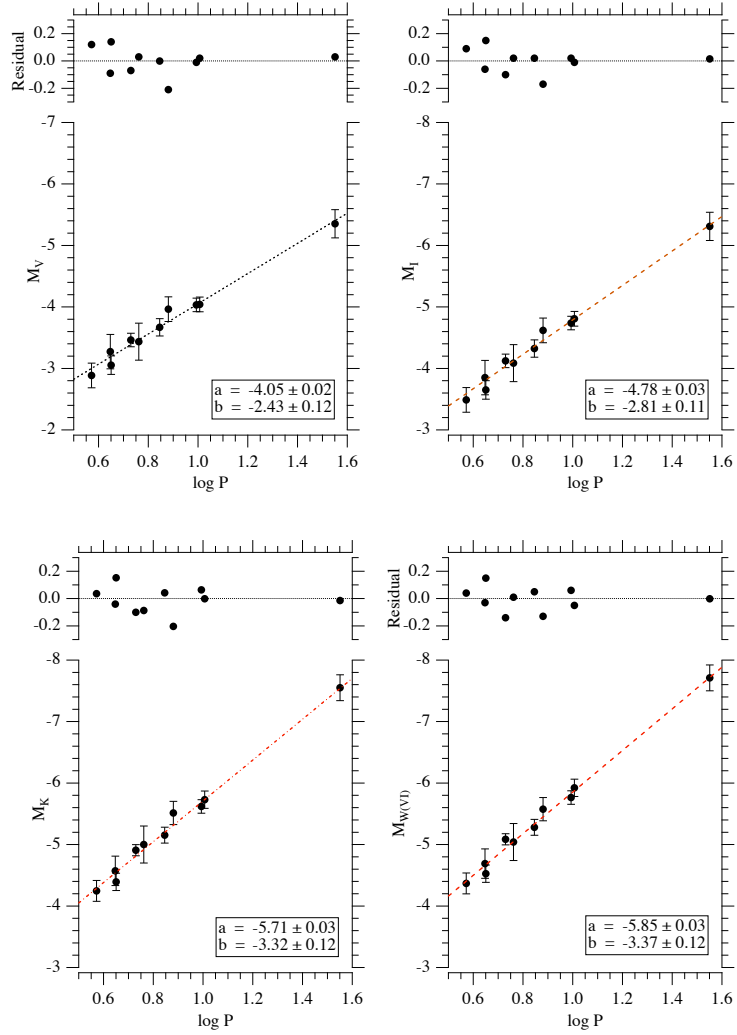}
\caption{Period-Luminosity relations for V, I, K, and Wesenheit W$_{VI}$, where W$_{VI}$ = $\langle$V$\rangle$-2.45($\langle$V$\rangle$-$\langle$I$\rangle$). Coefficients are for M $= a +b*(logP-1)$. Errors are $1\sigma$.} \label{fig-5}
\end{figure}

\begin{figure}
\epsscale{0.75}
\plotone{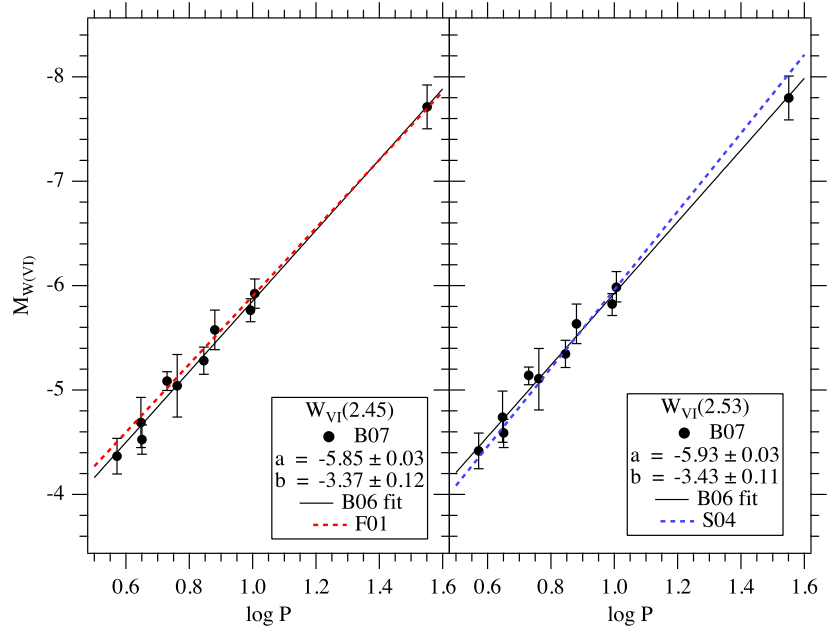}
\caption{M$_{W(VI)}$ Period-Luminosity relations for the present work compared to those adopted by \cite{Freed01} (left) and \cite{San04} (right). For the Freedman comparison we use 2.45 as our W$_{VI}$ color coefficient. For the Sandage comparison we recalculate our W$_{VI}$ using a 2.52 coefficient. While the Sandage \etal PLR agrees better with the bulk of our Cepheids, the Freedman \etal PLR better matches ours for logP$>$1.  Most H$_0$ Key Project extragalactic Cepheids are long-period. Hence our new PLR has minimal impact on the  Freedman \etal H$_0$ value. If adopted, our PLR would tend to increase the Sandage \etal (2006) H$_0$ value.} \label{fig-6}
\end{figure}

\begin{figure}
\epsscale{0.75}
\plotone{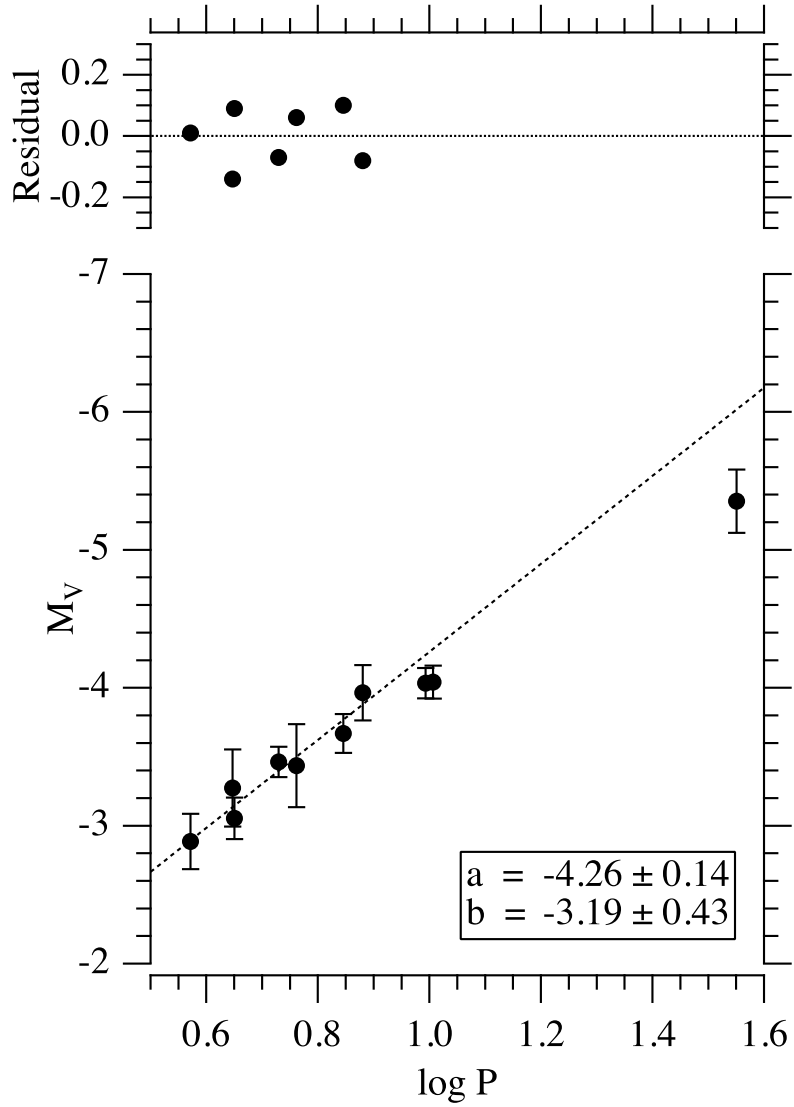}
\caption{Period-Luminosity relation for M$_V$  where we fit the seven Cepheids in our sample with logP$ < 1$. This fit excludes \lC~, $\zeta$ Gem, and $\beta$ Dor which has logP = 0.9931. Coefficients are for M $= a +b(logP-1.0)$. The slopes differ by $\sim2\sigma$, comparing the logP $<$ 1 slope against the entire range (Figure~\ref{fig-5}). Our parallaxes provide only weak evidence for a break at logP = 1.} \label{fig-PLRlt1}
\end{figure}

6.3.2.1 - LMC V-band PLR Non-linearity: 
 \cite{Nge05,Nge06, NgeK06} offer further evidence for a possible change of slope of the V-band PLR at logP$\sim$1 for the Cepheids in the LMC first noted by Sandage \etal. However, they find no such slope change in any reddening-free Wesenheit magnitude.  {\bf DC} With our small scatter V-band PLR do Galactic Cepheids exhibit a similar non-linearity? Fitting only those seven Cepheids in Figure~\ref{fig-5} with logP $<$ 1 for the V-band PLR results in a slope with a significant error.  Our fit to that period-restricted subset is shown in Figure~\ref{fig-PLRlt1}, and demonstrates that our sample, containing only one Cepheid with logP well in excess of unity, is too small to offer solid evidence for a V-band PLR slope change similar to that found in the LMC by Ngeow \etal. We note that \lC~  lies only $\sim2\sigma$ below the relationship line. We also note that we can obtain a V-band PLR slope only $1\sigma$ different from the full sample by retaining $\zeta$ Gem, and $\beta$ Dor in the sample. Suspected non-linearity in V rests entirely on \lC. Because the PLR in V is a collapsed period-luminosity-color relation, 
it has a finite width in V (e.g.,  Caldwell \& Coulson 1986\nocite{Cal86}).  The 
location of \lC~below the PLR fitted to shorter period stars 
could be a result of this finite width.
 
\begin{figure}
\epsscale{1.0}
\plotone{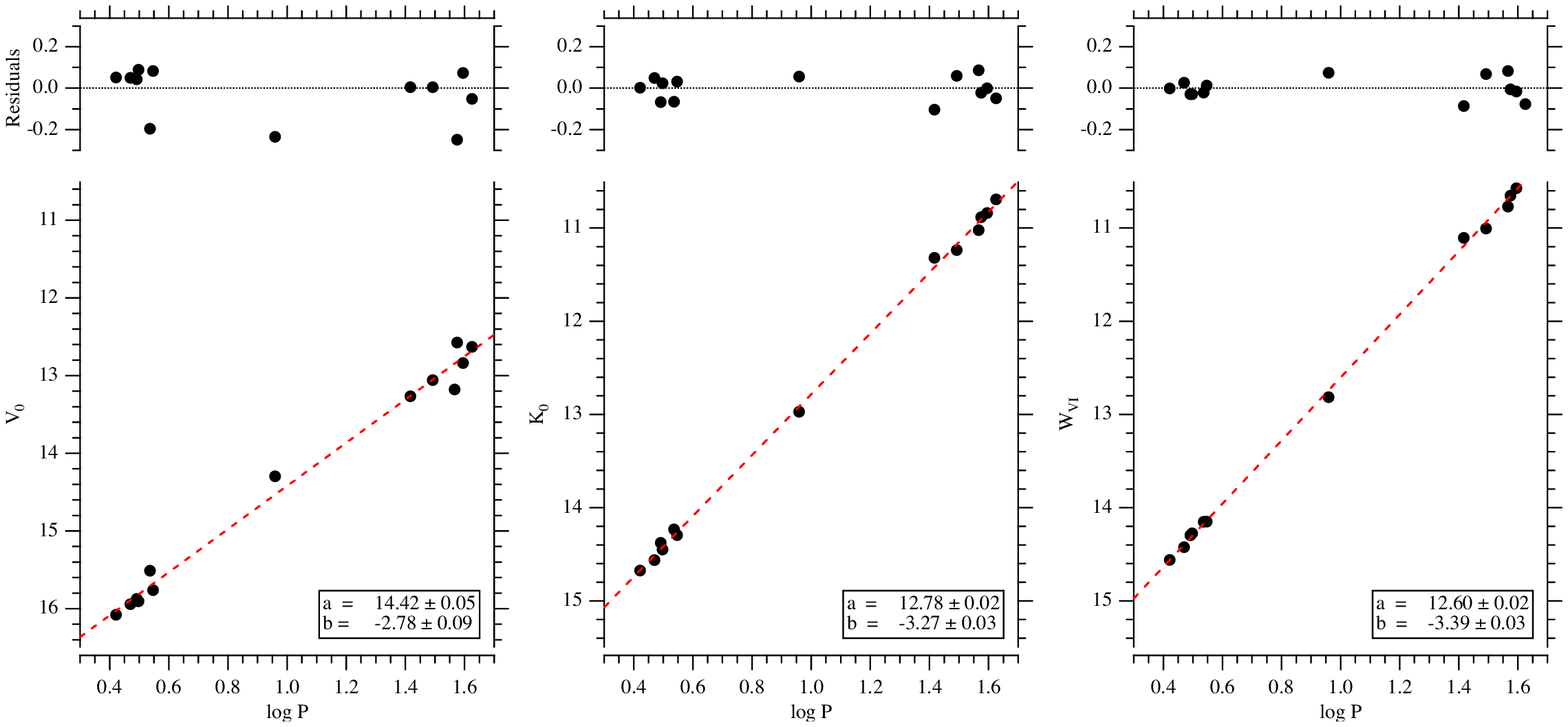}
\caption{Period-Luminosity relations for absorption-corrected V,  K, and W$_{VI}$  for 13 LMC Cepheids from \citet{Gie05}.  The K-band magnitudes have been transformed to the CIT system.  Coefficients are for M $= a +b(logP-1.0)$. The K-band and W$_{VI}$ slopes are in better agreement with the Galactic Cepheid results  than the V-band slope (Figure~\ref{fig-5}). } \label{fig-7}
\end{figure}

6.3.2.2 - Gieren \etal (2005):
contains  absolute magnitude information for thirteen LMC Cepheids derived using surface brightness methods. In addition to high-precision photometry, ten of the thirteen have metallicity measures ($\langle$[Fe/H]$\rangle$=-0.46). Because \cite{Gie05} contained none of the observed  intensity-averaged apparent magnitudes, J. Storm, a co-author on that paper, kindly supplied their   $\langle$V$\rangle$,  $\langle$I$\rangle$, and $\langle$K$\rangle$ for these selected LMC Cepheids. 
The $\langle$V$\rangle$, $\langle$I$\rangle$ and $\langle$K$\rangle$  were not corrected for LMC tilt, but are corrected for absorption (A$_V$= 3.1E(B-V),  A$_K$= 0.34E(B-V)), using  the E(B-V) from \citet{Gie05}).  The V and K  PLR, along with a Wesenheit W$_{VI}$ PLR (W$_{VI}$=V-2.45(V-I)), are shown in Figure~\ref{fig-7}. \\ {\bf DC} Comparing  the Gieren et al. LMC PLR Figure~\ref{fig-7} PLR with our Galactic PLR in Figure~\ref{fig-5} we note satisfactory agreement in the slopes for K and W$_{VI}$. The Gieren K data are a selected subset of the Persson \etal data discussed next.
The disagreement in V may be attributed to instability width \citep{Cal86} and the placement of \lC~ within that strip.

6.3.2.3 - Persson \etal (2004) K-band:
present the most extensive
infrared photometry (CIT system) of LMC Cepheids. An infrared PLR is of interest because it is less sensitive
to uncertainties in interstellar extinction and the intrinsic
width of the relation is likely to be small.  \\{\bf DC} Their PLR can be written
\begin{equation}
 K = -3.261(logP - 1) + 12.775
\end{equation}
The standard error of the slope is 0.04. This slope
is not significantly different from the one we determine from
our galactic stars ($-3.37\pm 0.09$, Table~\ref{tbl-11}). From Table~\ref{tbl-14} we
find a K-band Galactic Cepheid zero-point -5.67$\pm$0.03 for a PLR with slope -3.26. This direct comparison yields a K-band LMC distance modulus of 18.45 $\pm$ 0.04, uncorrected for metallicity effects.\\
 {\bf RP} Using the method of reduced parallaxes  and the Persson \etal slope,
our parallaxes yield M(K) = -3.261(logP -1) -5.678 ($\pm$ 0.033). 
We estimate that the uncertainty in the Persson \etal
LMC zero-point (in the shorter period range where our
Galactic stars are), is about 0.03 mag. Combined with our K-band PLR
and its
uncertainty the Persson \etal relationship  yields an LMC modulus of
$18.45 \pm 0.04$ without metallicity correction. We return to the metallicity issue in Section~\ref{LMCM}.

\begin{figure}
\epsscale{0.75}
\plotone{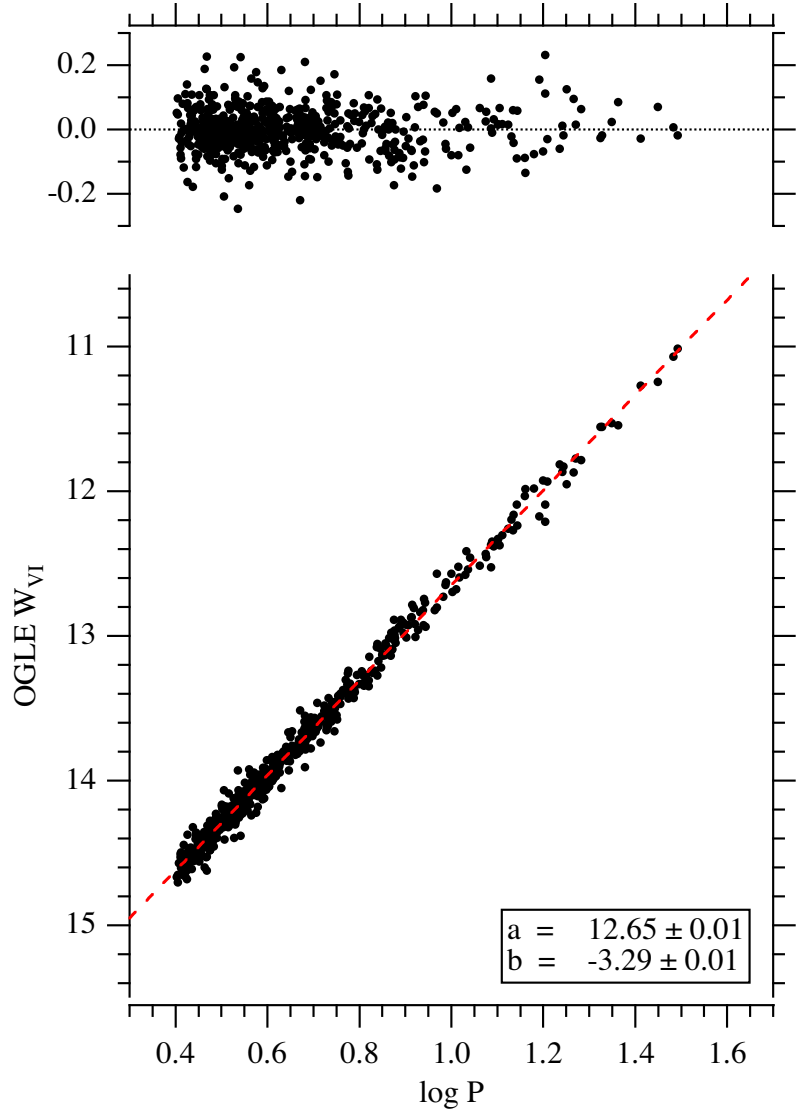}
\caption{W$_{VI}$ Period-Luminosity relation for 581 LMC Cepheids from OGLE.  Coefficients are for M $= a +b(logP-1.0)$ and W$_{VI}$ = $\langle$V$\rangle$-2.45($\langle$V$\rangle$-$\langle$I$\rangle$). } \label{fig-STO}
\end{figure}

6.3.2.4 - OGLE (1999) W$_{VI}$:
has produced the largest amount of LMC Cepheid photometry. In Figure~\ref{fig-STO} we plot an apparent  W$_{VI}$ PLR for 581 Cepheids in the LMC. These data were carefully preened, selecting only Cepheids with normal light curves and amplitudes \citep{Nge05}. They were kindly communicated by G. Tammann, and provide the  highly precise slope and zero point listed in Table~\ref{tbl-14}. \\ {\bf DC} Direct comparison of the W$_{VI}$ and M$_{W(VI)}$ zero-points from Figures~\ref{fig-STO} and~\ref{fig-5} yields an LMC distance modulus 18.49 $\pm$0.03 with no metallicity corrections. Constraining the slope to the OGLE value results in the LMC distance modulus 18.51 $\pm$0.04 listed in Table~\ref{tbl-14}.\\
{\bf RP} These data, when fit to  the relation
\beq
W(VI) = V- 2.45(V-I) = b(logP -1) + A  
\eeq
yield slope, b = -3.29$\pm$0.01,  which agrees within the uncertainties with the slope derived
from our Galactic stars.  For these we obtained b = -3.37 $\pm$ 0.12
and - 3.30 $\pm$ 0.12 in two slightly different solutions (see
Section 6.1 and  Table \ref{tbl-11}). Thus we find no evidence for a difference in
slope for the two galaxies. Nevertheless, because such a difference
has been suggested (Sandage \etal 2004) it is desirable to compare Galactic and LMC Cepheids in the same period range. In the case of the Galactic stars this omits
\lC.  We therefore solve equation (11) for Cepheids with log P
in the range 0.5 to 1.1. The OGLE LMC data then yield b = - 3.31 $\pm$ 0.03
and A = 12.64 $\pm$ 0.01, not different from the values for the
whole sample. Adopting a slope b = -3.31 we find a = -5.85 $\pm$ 0.04 in the equation
\beq
M_{W(VI)} = -3.31(log P -1) + a  
\eeq
from our Cepheids. This leads to an LMC modulus of A-a = 18.50 $\pm$ 0.04,
uncorrected for metallicity effects in W$_{VI}$.

To summarize this section our PLR can be used to obtain LMC distance moduli by comparing ({\bf DC}) our absolute zero-points with apparent magnitude zero-points from OGLE, Perrson \etal, and Gieren \etal. Constraining the slopes of the PLR in Figure~\ref{fig-5} to those determined from the Gieren and OGLE PLR in Figures~\ref{fig-7} and \ref{fig-STO} yields the zero points, a(DC), found in Table~\ref{tbl-14}. Comparing the zero-points for all four PLR we find LMC distance moduli 18.45 $\pm$0.04 for the K band, 18.51$\pm$0.04 for W$_{VI}$,  18.52 $\pm$0.06 for the V band, and 18.49 $\pm$0.03 for the OGLE W$_{VI}$. {\bf RP} yields the zero-points a(RP) in Table~\ref{tbl-14} and LMC distance moduli 18.50$\pm$0.04 for W$_{VI}$ and 18.42 $\pm$ 0.04 for K. These moduli remain uncorrected for possible metallicity effects.

\subsection{Extragalactic Distances: Applying Our PLR}
 In this section we apply our PLR to the LMC and NGC 4258, comparing our derived distances with those from other investigators. In the case of the LMC we briefly describe our adopted metallicity corrections.
\subsubsection{Metallicity Corrections and LMC Distance Modulus} \label{LMCM}
 Note that none of the LMC distance moduli derived  above (Table~\ref{tbl-14}) have  metallicity corrections applied.  \cite{Mac06} demonstrate that a metallicity correction is necessary by comparing  metal-rich Cepheids with  metal-poor Cepheids in NGC 4258.  With a previously measured  [O/H] metallicity gradient \citep{Zar94} Macri \etal find a Cepheid metallicity correction in W$_{VI}$, $\gamma = -0.29 \pm 0.09_r \pm 0.05_s$ magnitude for 1 dex in metallicity, where r and s subscripts signify random and systematic. This value is similar
to an earlier W$_{VI}$ metallicity correction (Kennicutt et al. 1998) derived from  Cepheids in  M101
(-0.24$\pm$ 0.16). Other less direct determinations (based for
instance on RGB tip distances and  Baade-Wesselink type
luminosities) are summarized by Macri \etal and are in agreement with these figures. Taking the weighted mean of the Kennicutt and Macri values
and using the difference in metallicity of LMC and Galactic Cepheids (-0.36 dex from means of the data in
Groenewegen \etal 2004 tables 3 and 4) we find a metallicity
correction of -0.10$\pm$ 0.03 magnitude with the Galactic Cepheids being
brighter. The LMC distance moduli in Table~\ref{tbl-14} from the Persson \etal K data in the previous subsection suggest that the metallicity correction may be smaller for K than for W$_{VI}$.

Returning to the issue of the true distance modulus to the LMC, our lowest error estimate  is derived from the the OGLE photometry (Section 6.3.2.4, OGLE:  m-M = 18.50$\pm$ 0.04). Combined with the estimated metallicity
correction (-0.10$\pm$ 0.03 magnitude) we obtain an LMC modulus of 18.40 $\pm$ 0.05.  \cite{Ben02a} lists 84 determinations complete through 2001 which can be compared with our new modulus. One recent determination is noteworthy for its lack of dependence on any metallicity corrections.  \cite{Fit03} derive 18.42$\pm$ 0.04, 
from eclipsing binaries, a modulus in excellent agreement with our new value.

\subsubsection{NGC 4258 Distance Modulus}
Using \HST~ photometry of NGC 4258 Cepheids, \cite{Mac06} have determined a distance modulus  relative to the LMC. They find  that the difference in distance moduli
of the LMC and NGC4258 is 10.88$\pm$ 0.06 mag. NGC 4258 has an independently measured distance obtained by comparing circumnuclear maser proper motions and radial velocities  \citep{Her99}. Macri \etal surveyed two fields in NGC 4258, one near the nucleus, the other in the outer regions of the galaxy.  Macri (private communication) has applied our W$_{VI}$ PLR (Figure~\ref{fig-5}) directly to N=85 inner field, solar metallicity Cepheids and finds m-M = 29.21 $\pm$ 0.02. The maser distance modulus is m-M = 29.29 $\pm$ 0.15. 
Our adopted LMC modulus, m-M= 18.40 $\pm$ 0.05 and the Macri differential modulus (LMC-NGC 4258) leads to a modulus of 29.28 $\pm$ 0.08 for NGC 4258, a value in even closer agreement with the maser-based distance. 

\section{Summary}
\begin{enumerate}
\item {\it HST} astrometry has now yielded absolute trigonometric parallaxes  for 10 Cepheid variables with an average $\sigma_\pi = 0.2 $ mas, or $\sigma_\pi/\pi = 8$\%. These parallaxes, along with precision photometry culled from the literature, Lutz-Kelker-Hanson bias corrections,  and reddening corrections derived from both the literature and our ancillary spectrophotometry, provide absolute magnitudes with which to form Period-Luminosity relations. At logP = 1, our zero-point errors are now at or below 0.03 magnitudes in all bandpasses.

\item Including perturbation orbits in our astrometry for W Sgr and FF Aql results in Cepheid orbit size and perturbation inclination. Assuming masses for the secondaries consistent with their known spectral type yields relatively low precision masses for these two Cepheids. We find for W Sgr and FF Aql, respectively, $M = 6.5 \pm 2 M_{\sun}$ and $M = 4.5 \pm 1 M_{\sun}$. The major contributor to the mass uncertainty remains the parallax for FF Aql and the secondary spectral type for W Sgr.

\item Comparing our parallax of T Vul with a parallax 
determined through the surface brightness technique for that 
Cepheid, we find agreement at the 1$\sigma$ level.  Alternately, 
if we demand that the surface brightness parallax be the same as 
our \HST~parallax, we determine a quasi-geometrical value of the 
radial velocity p-factor, $p=1.19\pm0.16$.  Our PLR in the V 
magnitude agrees within 1$\sigma$ in slope and zero point with the 
Galactic PLR relation based on the Bayesian surface brightness PLR 
of Barnes et al. 

\item Comparing  our W$_{VI}$ zero-points with those adopted by the Freedman and Sandage H$_0$ projects, we find better overall zero-point agreement with Sandage. However, the PLR adopted by Freedman \etal agrees with ours at longer periods. Given that most of the Cepheids observed in external galaxies are long-period,  there may be little effect on the Freedman \etal H$_0$ value. Adopting our PLR would increase the Sandage \etal H$_0$ value.

\item Comparing our V, K, and W$_{VI}$ PLR with LMC PLR  we find slope agreement for K and W$_{VI}$ within the errors. The disagreement in V may be attributed to instability width and the placement of \lC~ within that strip. Comparing (both direct comparisons and via reduced parallaxes) zero-points yields a W$_{VI}$ LMC distance modulus. (m-M) = 18.50 $\pm 0.04$, without any metallicity correction. Adopting a metallicity correction -0.10$\pm$ 0.03 magnitude between Galactic and LMC Cepheids (with Galactic being brighter), we find a true LMC distance modulus (m-M)$_0$ = 18.40 $\pm$ 0.05.

\item Applying our PLR directly to Cepheids in NGC 4258 provides a distance modulus, m-M = 29.21 $\pm$ 0.02, in good agreement with the maser distance modulus, m-M = 29.29 $\pm$ 0.15. From a metallicity-corrected LMC distance modulus and the Macri \etal difference in distance moduli we obtain m-M = 29.28 $\pm$ 0.08.  

\end{enumerate}

\acknowledgments

We thank Lisa Crause (SAAO) for obtaining and reducing some of the BVRI CCD photometry of southern Cepheid reference star fields.
Support for this work was provided by NASA through grants GO-09879 and GO-10106  from the Space Telescope Science Institute, which is operated
by AURA, Inc., under NASA contract NAS5-26555. This paper uses observations made at the South African Astronomical Observatory (SAAO), and observations obtained with the Apache Point Observatory 3.5-meter telescope, which is owned and operated by the Astrophysical Research Consortium. This paper uses observations made at the Kitt Peak National Observatory 4m telescope. KPNO is also operated by AURA. This publication makes use of data products from the Two Micron All Sky Survey, which is a joint project of the University of Massachusetts and the Infrared Processing and Analysis Center/California Institute of Technology, funded by NASA and the NSF. This research has made use of the SIMBAD database, operated at CDS, Strasbourg, France; the NASA/IPAC Extragalactic Database (NED) which is operated by JPL, California Institute of Technology, under contract with the NASA;  and NASA's Astrophysics Data System Abstract Service.  This material is based upon work by TGB while serving at the National Science Foundation. Any opinions, findings, and conclusions or recommendations expressed in this material are those of the authors and do not necessarily reflect the views of the National Science 
Foundation. We thank Jesper Storm for kindly providing the LMC Cepheid apparent magnitudes discussed in Section 6.3.2.2. We thank G. Tammann for access to the preened OGLE photometry in Section 6.3.2.4. We thank Bill Jefferys for discussions regarding Bayes Theorem and the LKH bias correction. We thank Lucas Macri for valuable discussions, and for applying our PLR to his NGC 4258 Cepheid sample. We thank an anonymous referee for constructive criticism.
\clearpage


%

\clearpage

\begin{deluxetable}{llllrrr}
\tablewidth{6in}
\tablecaption{Log of Observations, Pulsational Phase, and Estimated B-V\label{tbl-LOO}}
\tablehead{\colhead{Set}&
\colhead{mJD}&\colhead{Phase}&\colhead{B-V\tablenotemark{a}} &\colhead{mJD}&\colhead{Phase}&\colhead{B-V}\\
}
\startdata
&\lC&&&$\zeta$ Gem&&\\
1&52816.56612&0.845&1.29&52917.92253&0.709&0.77\\
2&52968.8092&0.128&1.14&52923.85887&0.294&0.95\\
3&52969.81053&0.156&1.17&53023.39599&0.098&0.81\\
4&53100.66518&0.837&1.30&53097.58775&0.407&0.96\\
5&53161.31783&0.544&1.47&53099.25408&0.571&0.89\\
6&53162.38365&0.574&1.47&53136.11753&0.202&0.90\\
7&53334.73308&0.422&1.42&53283.55272&0.724&0.76\\
8&53335.79732&0.452&1.44&53288.02009&0.165&0.87\\
9&53465.02929&0.087&1.09&53390.30801&0.240&0.92\\
10&53525.78567&0.796&1.35&53460.40544&0.145&0.85\\
11&53527.31753&0.839&1.30&53464.40293&0.539&0.91\\
&$\beta Dor$&&&W Sgr&&\\
1&52897.7079&0.494&0.97&52823.5874&0.185&0.66\\
2&52897.7742&0.501&0.97&52905.61983&0.986&0.50\\
3&52953.3841&0.151&0.75&52910.28674&0.600&0.94\\
4&53077.2435&0.735&0.82&52940.08889&0.524&0.91\\
5&53080.1724&0.033&0.67&53081.02535&0.081&0.59\\
6&53127.1734&0.808&0.75&53086.62384&0.818&0.86\\
7&53259.8937&0.293&0.89&53272.31547&0.268&0.73\\
8&53263.1581&0.624&0.92&53276.11681&0.768&0.91\\
9&53316.8711&0.082&0.69&53306.17434&0.726&0.94\\
10&53439.2811&0.519&0.97&53447.3796&0.318&0.77\\
11&53445.1417&0.114&0.72&53451.77831&0.897&0.74\\
&X Sgr&&&Y Sgr&&\\
1&52905.686&0.576&0.90&52907.28649&0.700&1.02\\
2&52910.34909&0.241&0.76&52913.28455&0.739&1.01\\
3&52937.01735&0.044&0.64&53052.09011&0.781&0.97\\
4&53080.08851&0.444&0.89&53087.62197&0.935&0.75\\
5&53084.95383&0.137&0.68&53093.88707&0.021&0.68\\
6&53170.0758&0.274&0.79&53157.27431&0.600&0.62\\
7&53272.24833&0.843&0.70&53273.45131&0.123&0.76\\
8&53275.1117&0.251&0.77&53279.2493&0.127&0.76\\
9&53305.23779&0.547&0.91&53416.92419&0.973&0.66\\
10&53445.70857&0.576&0.90&53453.84083&0.368&0.94\\
11&53449.77493&0.155&0.69&53458.17128&0.118&0.76\\
&FF Aql&&&T Vul&&\\
1&52826.66128&0.433&0.85&52895.35431&0.018&0.46\\
2&52919.09143&0.107&0.74&52956.16164&0.727&0.77\\
3&52924.35973&0.285&0.81&52960.16233&0.629&0.80\\
4&53047.03016&0.723&0.81&53080.96436&0.865&0.64\\
5&53102.0262&0.023&0.70&53137.95919&0.715&0.77\\
6&53106.02418&0.918&0.73&53143.82222&0.037&0.48\\
7&53285.31579&0.019&0.70&53322.17372&0.247&0.66\\
8&53290.24664&0.122&0.75&53326.3049&0.178&0.61\\
9&53416.9962&0.472&0.85&53444.84837&0.905&0.58\\
10&53469.90158&0.305&0.82&53502.01713&0.794&0.72\\
11&53471.56806&0.678&0.82&53507.07994&0.935&0.53\\
&RT Aur&&&&&\\
1&52910.45334&0.719&0.78&&&\\
2&52915.98709&0.203&0.55&&&\\
3&52996.65789&0.841&0.70&&&\\
4&53081.79189&0.676&0.79&&&\\
5&53085.45754&0.660&0.79&&&\\
6&53129.1859&0.389&0.68&&&\\
7&53278.95104&0.560&0.77&&&\\
8&53281.95338&0.365&0.66&&&\\
9&53371.84211&0.475&0.73&&&\\
10&53446.5509&0.514&0.75&&&\\
11&53453.41092&0.354&0.66&&&\\
\enddata
\tablenotetext{a}{B-V estimated from phased light curve.}
\end{deluxetable}

\begin{deluxetable}{l l l l l l l l l l}
\tablewidth{0in}
\tablecaption{Target Cepheid Apparent Properties \label{tbl-AP}} 
\tablehead{\colhead{ID}&
\colhead{P(days)} &
\colhead{log P} &
\colhead{$\langle$V$\rangle$} &
\colhead{$\langle$I$\rangle$\tablenotemark{a}} &
\colhead{$\langle$K$\rangle$\tablenotemark{b}} &
\colhead{$\langle$B-V$\rangle$} &
\colhead{E(B-V)} &
\colhead{A$_V$} &
\colhead{A$_K$}
}
\startdata
\lC&35.551341&1.5509&3.732&2.557&1.071&1.299&0.17&0.52&0.06\\
$\zeta$ Gem&10.15073&1.0065&3.911&3.085&2.097&0.798&0.018&0.06&0.01\\
$\beta$ Dor&9.842425&0.9931&3.751&2.943&1.944&0.807&0.044&0.25&0.03\\
W Sgr&7.594904&0.8805&4.667&3.862&2.796&0.746&0.111&0.37&0.04\\
X Sgr&7.012877&0.8459&4.556&3.661&2.557&0.739&0.197&0.58&0.07\\
Y Sgr&5.77338&0.7614&5.743&4.814&3.582&0.856&0.205&0.67&0.07\\
$\delta$ Cep&5.36627&0.7297&3.960&3.204&2.310&0.657&0.092&0.23&0.03\\
FF Aql&4.470916&0.6504&5.372&4.510&3.465&0.756&0.224&0.64&0.08\\
T Vul&4.435462&0.6469&5.752&5.052&4.187&0.635&0.064&0.34&0.02\\
RT Aur&3.72819&0.5715&5.464&4.778&3.925&0.595&0.051&0.20&0.02\\
\enddata
\tablenotetext{a}{Cousins I}
\tablenotetext{b}{CIT K}
\end{deluxetable}
\begin{deluxetable}{l l l l l l l l l}
\tablewidth{0in}
\tablecaption{\lC~ Reference Stars: Visible and Near-IR Photometry \label{tbl-IR}}
\tablehead{\colhead{ID}&
\colhead{FGS ID}&
\colhead{V} &
\colhead{B-V} &
\colhead{U-B} &
\colhead{V-I\tablenotemark{a}} &
\colhead{K\tablenotemark{a}} &
\colhead{J-K\tablenotemark{a}} &
\colhead{V-K\tablenotemark{a}} }
\startdata
4273957\tablenotemark{b}&2&14.32&0.71&0.30&0.89&12.52&0.30&1.80\\
4273905\tablenotemark{b}&4&13.53&0.95&0.63&1.13&11.20&0.62&2.33\\
2M\tablenotemark{c}&5&13.23&1.18&1.08&1.33&10.29&0.78&2.95\\
4066585\tablenotemark{b}&8&10.77&1.58&1.95&1.87&6.56&1.10&4.22\\
4066439\tablenotemark{b}&9&13.49&0.57&0.08&0.72&&&\\
4066556\tablenotemark{b}&10&13.01&0.60&-0.04&0.80&11.48&0.34&1.53\\
\enddata
\tablenotetext{a}{Cousins I, Bessell/Brett JHK}
\tablenotetext{b}{ID from UCAC2 catalog}
\tablenotetext{c}{ID= 09454541-6230004 from 2MASS catalog}
\end{deluxetable}

\begin{center}
\begin{deluxetable}{rccccccc}
\tablewidth{0in}
\tablecaption{\lC~ Field A$_V$ from Reference Star Spectrophotometry  \label{tbl-AV}}
\tablehead{  \colhead{ID}&
\colhead{A$_V$(B-V)}&   \colhead{A$_V$(V-I)}&  \colhead{A$_V$(V-K)} &  \colhead{A$_V$(J-K)}&  \colhead{A$_V$(U-B)}&
\colhead{$\langle$A$_V\rangle$}&\colhead{SpT}
}
\startdata
2&0.4&0.6&0.4&-0.3&0.6&0.33$\pm$0.19&G0V\\
4&0.0&0.5&0.2&0.2&-0.2&0.14 0.13&G8III\\
5&0.5&0.8&0.7&0.9&0.7&0.72 0.07&K0III\\
8&0.6&1.1&1.0&1.3&1.2&1.02 0.14&K4III\\
9&0.5&0.6&0.0&0.0&0.2&0.45 0.12&F3V\\
10&0.6&0.8&0.5&0.5&-0.2&0.46 0.18&F4V\\
\\
$\langle$A$_V\rangle$&0.45&0.73&0.55&0.52&0.38&{\bf 0.52}&\\
$\pm$&0.10&0.09&0.15&0.31&0.24&{\bf 0.06}&\\
\enddata
\end{deluxetable}
\end{center}

\begin{deluxetable}{lllllrl}
\tablewidth{0in}
\tablecaption{\lC~ Astrometric Reference Star 
Spectrophotometric Parallaxes \label{tbl-SPP}}
\tablehead{
\colhead{ID}& \colhead{V} &\colhead{Sp. T.}&
 \colhead{M$_V$} & \colhead{A$_V$} &\colhead{m-M}& 
\colhead{$\pi_{abs}$(mas)}
} 
\startdata
2&14.32&G0V&4.4&0.33&9.89$\pm$0.5&1.2$\pm$0.3\\
4&13.53&G8III&0.9&0.14&12.63 0.7& 0.3  0.1\\
5&13.23&K0III&0.7&0.72&12.48 0.5& 0.4  0.1\\
8&10.77&K4III&-0.1&1.02&10.72 0.5& 1.1  0.3\\
9&13.49&F3V&3.2&0.45&10.24 0.5& 1.1 0.3\\
10&13.01&F4V&3.3&0.46&9.67 0.5& 1.4  0.3\\
\enddata 
\end{deluxetable}

\begin{deluxetable}{llrr}
\tablewidth{0in}
\tablecaption{\lC~ and Reference Star Relative Positions \tablenotemark{a} \label{tbl-POS}}
\tablehead{\colhead{FGS ID}&
\colhead{V} &
\colhead{$\xi$ \tablenotemark{b}} &
\colhead{$\eta$ \tablenotemark{b}} 
}
\startdata
\lC&3.72&51.0107$\pm$0.0005&29.4377$\pm$0.0003\\
2&14.32&20.4338 0.0006&133.6164 0.0005\\
4&13.53&-61.6015 0.0007&64.6358 0.0007\\
5&13.23&262.9969 0.0008&57.4027 0.0006\\
8&10.77&199.6315 0.0006&36.9602 0.0004\\
9\tablenotemark{c}&13.49&0.0000 0.0006&0.0000 0.0005\\
10&13.01&161.3411 0.0006&-31.8256 0.0005\\
\enddata
\tablenotetext{a}{epoch 2004.431}
\tablenotetext{b}{$\xi$ and $\eta$ are relative positions in arcseconds
}
\tablenotetext{c}{RA = 09 45 07.44	 , Dec = -62 30 57.9, J2000, epoch 2004.431}
\end{deluxetable}

\begin{center}
\begin{deluxetable}{lll}
\tablewidth{3in}
\tablecaption{\lC~ and Reference Star Relative Proper Motions\label{tbl-PM}}
\tablehead{\colhead{ID}&
\colhead{$\mu_x$\tablenotemark{a}} &
\colhead{$\mu_y$\tablenotemark{a}}
 }
\startdata
\lC&-0.0126$\pm$0.0003&0.0085$\pm$0.0004\\
2&-0.0098 0.0010&0.0094 0.0011\\
4&-0.0009 0.0012&0.0096 0.0013\\
5&-0.0073 0.0017&0.0068 0.0016\\
8&-0.0056 0.0012&0.0040 0.0011\\
9&-0.0106 0.0009&0.0064 0.0008\\
10&-0.0066 0.0010&0.0055 0.0011\\
\enddata
\tablenotetext{a}{$\mu_x$ and $\mu_y$ are relative motions in arcsec
yr$^{-1}$ }
\end{deluxetable}
\end{center}

\begin{center}
\begin{deluxetable}{lllr}
\tablewidth{4in}
\tablecaption{\lC~ and Reference Star Parallaxes and Transverse Velocities    \label{tbl-PiVt}}
\tablehead{
\colhead{ID}&
\colhead{$\mu$\tablenotemark{a,b}} &
\colhead{$\pi_{abs}$\tablenotemark{b}} &
\colhead{V$_t$\tablenotemark{c}} 
\\
\colhead{}&
\colhead{(mas yr$^{-1}$)}&
\colhead{(mas)}&
\colhead{(\kms)}
}
\startdata
\lC\tablenotemark{d}& 15.2&2.01$\pm$0.20&36$\pm$4\\
2& 13.5&1.33 0.14&48 6\\
4& 9.6&0.32 0.04&144 75\\
5&10.0&0.45 0.05&105 14\\
8&6.8&1.19 0.10&27 4\\
9&12.4&0.73 0.22&80 25\\
10& 8.6&1.45 0.11&28 3\\
\enddata
\tablenotetext{a}{ $\mu = (\mu_x^2 +\mu_y^2)^{1/2}$ from  Table~\ref{tbl-PM}}
\tablenotetext{b}{Final from modeling with equations 2 -- 5}
\tablenotetext{c}{V$_t = 4.74\times \mu/\pi_{abs}$}
\tablenotetext{d}{Modeled with equations 2 -- 5, constraining D=-B and E=A}
\end{deluxetable}
\end{center}

\begin{deluxetable}{llll}
\tablewidth{4in}
\tablecaption{Astrometric Reference Star 
Final Parallaxes \label{tbl-allP}}
\tablehead{
\colhead{ID\tablenotemark{1}}& \colhead{V} &\colhead{Sp. T.}&
\colhead{$\pi_{abs}$(mas)}
} 
\startdata
LC-2&14.29&G0V&1.3$\pm$0.1\\
LC-4&13.53&G8III&0.3 0.1\\
LC-5&13.18&K0III&0.4 0.1\\
LC-8&10.62&K4III&1.2 0.1\\
LC-9&13.44&F3V&0.7 0.2\\
LC-10&12.97&F4V&1.5 0.1\\
\\
ZG-2&13.78&G8III&0.3 0.1\\
ZG-3&11.47&F3.5V&2.2 0.1\\
ZG-5&12.36&F6V&1.9 0.1\\
ZG-8&7.55&G3V&27.2 0.2\\
ZG-10&14.25&F5V&0.8 0.1\\
ZG-11&12.56&K0III&0.5 0.1\\
\\
BD-2&15.84&A0III&0.1 0.1\\
BD-3&13.26&F5V&1.3 0.1\\
BD-4&15.79&G3V &0.6 0.1\\
BD-5&14.70&G9V&1.7 0.1\\
BD-6&15.28&K0III&0.1 0.1\\
BD-7&15.29&G5V&0.9 0.1\\
BD-8&16.42&K0V?&0.7 0.1\\
\\
WS-4&11.25&F1V&2.5 0.1\\
WS-5&13.25&K0III&0.9 0.2\\
WS-7&12.8&K0III&0.6 0.1\\
WS-9&14.17&F8V&1.5 0.1\\
WS-10&13.7&M0III&0.3 0.1\\
WS-11&14.1&F2III&0.3 0.1\\
\\
XS-2&14.00&K0III&0.5 0.1\\
XS-3&13.10&B7V&0.5 0.1\\
XS-4&13.62&A1III&0.5 0.1\\
XS-5&12.56&K0III&0.9 0.1\\
XS-6&13.04&F5V&1.9 0.1\\
XS-7&12.56&F3V&1.9 0.1\\
XS-8&13.98&A1V&0.7 0.1\\
\\
YS-2&10.37&A5V&2.2 0.3\\
YS-3&12.41&A5V&1.0 0.1\\
YS-4&13.36&K0IV&1.6 0.2\\
YS-7&11.18&F0V&2.2 0.2\\
YS-9&14.92&K7V&4.9 0.5\\
YS-10&12.83&G9III&0.5 0.1\\
\\
FF-2&14.17&K2III&0.3 0.1\\
FF-3&14.16&K3V&3.6 0.2\\
FF-4&13.68&K3V&4.0 0.2\\
FF-5&14.93&G7V&1.6 0.1\\
FF-6&15.1&F2V&0.6 0.1\\
FF-7&15.29&K2III&0.2 0.1\\
\\
TV-2&13.79&K0III&0.4 0.1\\
TV-3&13.31&G3V&2.1 0.2\\
TV-4&14.29&K1IV&0.7 0.1\\
TV-5&13.26&G0V&1.5 0.2\\
TV-6&11.69&K1.5III&0.8 0.1\\
TV-7&14.48&K0IV&0.6 0.1\\
TV-8&12.60&K3III&0.6 0.1\\
\\
RT-4&13.87&K2V&2.7 0.2\\
RT-5&13.26&K0III&0.4 0.1\\
RT-6&11.37&G2V&3.0 0.3\\
RT-7&11.47&F3V&2.4 0.2\\
RT-8&13.90&F3V&0.9 0.1\\
RT-9&14.93&G5III&0.2 0.1\\
\enddata 
\tablenotetext{1}{ ~LC-2 = \lC, Ref-2; ZG = $\zeta$ Gem; BD = $beta$ Dor; WS = W Sgr; XS = X Sgr; YS = Y Sgr; FF = FF Aql; TV = T Vul; RT = RT Aur}
\end{deluxetable}

\begin{center}
\begin{deluxetable}{lll}
\tablecaption{Binary Cepheid Orbits and Masses \label{tbl-10}}
\tablewidth{0in}
\tablehead{\colhead{Parameter} &  \colhead{W Sgr}&  \colhead{FF Aql}}
\startdata
$\alpha$(mas)&2.67 $\pm$ 0.2&3.36 $\pm$ 0.4\\
P(days)& 1582 $\pm$ 3&1434 $\pm$ 1\\
P(years)& 4.33 $\pm$ 0.01& 3.93 $\pm$ 0.01\\
T$_0$ & 2004.16 $\pm$ 0.01& 2003.29 $\pm$ 0.04\\
e& 0.41 $\pm$ 0.02& 0.09 $\pm$ 0.01\\
i& 7\fdg0 $\pm$ 0\fdg8& 33\arcdeg $\pm$ 5\arcdeg \\
$\Omega$&68\fdg4 $\pm$ 4\fdg0& 61.3\arcdeg $\pm$ 9\arcdeg\\
$\omega$&328\fdg0 $\pm$ 1\fdg3& 327\arcdeg $\pm$ 4\arcdeg\\
\\
Secondary Sp. T.&A5V--F5V\tablenotemark{1} &F1V\tablenotemark{2}\\
Secondary Mass &2.0--1.4 &1.6\\
$a$ (mas)& 12.9 $\pm$ 0.3& 12.8 $\pm$ 0.9\\
$a$ (AU) & 5.67 $\pm$ 0.13 & 4.54 $\pm$ 0.14 \\
f& 0.207 $\pm$ 0.017& 0.263 $\pm$ 0.031\\
Cepheid Mass ($M_{\sun}$)&6.5 $\pm$ 2 &4.5 $\pm$ 1\\
\enddata
\tablenotetext{1}{Range from N. Evans (private communication)}
\tablenotetext{2}{\cite{Eva90}}
\end{deluxetable}
\end{center}

\begin{deluxetable}{rlllll}
\tablecaption{Cepheid Parallaxes, Proper Motions, and Absolute Magnitudes\label{tbl-SUM}} 
\tablewidth{0pt}
\tablehead{
\colhead{Parameter} & \colhead{} &\colhead{}&\colhead{Cepheid}&\colhead{}&\colhead{}
}
\startdata
&\lC&$\zeta$ Gem&$\beta$ Dor&W Sgr&X Sgr\\
Duration (y)&1.95&1.50&1.50&1.71&1.49\\
Ref stars (\#)&6&6&5&6&7\\
Ref $\langle$V$\rangle$&13.00&12.03&14.95&13.04&13.28\\
Ref $\langle$B-V$\rangle$&0.92&0.69&0.77&1.35&0.98\\
$\pi_{abs}$ (mas)&\lCpi&\zGpi&\bDpi&\WSpi& \XSpi \\
$\mu$ (mas y$^{-1}$)&15.2$\pm$0.5&6.2$\pm$0.5&12.7$\pm$0.8&6.6$\pm$0.4&10.0$\pm$1.2\\
P.A. (\arcdeg)&304$\pm$2&272$\pm$5&10.4$\pm$0.6&134$\pm$8&193$\pm$3\\
Av&0.52$\pm$0.06&0.06$\pm$0.03&0.25$\pm$0.05&0.37$\pm$0.03&0.58$\pm$0.1\\
LKH Corr&-0.08&-0.03&-0.02&-0.06&-0.03\\
(m-M)$_0$&8.56&7.81&7.50&8.31&7.64\\
M$_V$&-5.35$\pm$0.22&-4.03$\pm$0.15&-4.05$\pm$0.11&-3.97$\pm$0.20&-3.68$\pm$0.17\\
M$_I$&-6.31$\pm$0.22&-4.80$\pm$0.15&-4.74$\pm$0.11&-4.62$\pm$0.20&-4.64$\pm$0.17\\
M$_K$&-7.55$\pm$0.21&-5.73$\pm$0.14&-5.62$\pm$0.11&-5.51$\pm$0.19&-5.15$\pm$0.13\\
M$_{W(VI)}$&-7.71$\pm$0.21&-5.92$\pm$0.14&-5.76$\pm$0.11&-5.58$\pm$0.19&-5.28$\pm$0.13\\
\hline
&Y Sgr&$\delta$ Cep&FF Aql&T Vul&RT Aur\\
Duration (y)&1.51&2.44&1.77&1.67&1.49\\
Ref stars (\#)&6&5&6&6&5\\
Ref $\langle$V$\rangle$&12.51&12.06&14.48&13.39&13.02\\
Ref $\langle$B-V$\rangle$&0.98&1.30&1.16&1.12&0.80\\
$\pi_{abs}$ (mas)&\YSpi&\dCpi&\FApi&\TVpi&\RApi \\
$\mu$ (mas y$^{-1}$)&7.0$\pm$0.8&17.4$\pm$0.7&7.9$\pm$0.8&7.1$\pm$0.3&15.0$\pm$0.4\\
P.A. (\arcdeg)&204$\pm$5&-73$\pm$3&144$\pm$11&141$\pm$6&179$\pm$3\\
Av&0.67$\pm$0.04&0.23$\pm$0.03&0.64$\pm$0.06&0.34$\pm$0.06&0.20$\pm$0.08\\
LKH Corr&-0.15&-0.01&-0.03&-0.12&-0.05\\
(m-M)$_0$&8.51&7.19&7.79&8.73&8.15\\
M$_V$&-3.42$\pm$0.30&-3.47$\pm$0.11&-3.05$\pm$0.15&-3.24$\pm$0.28&-2.90$\pm$0.18\\
M$_I$&-4.12$\pm$0.30&-4.14$\pm$0.11&-3.65$\pm$0.27&-3.79$\pm$0.28&-3.47$\pm$0.18\\
M$_K$&-5.00$\pm$0.30&-4.91$\pm$0.09&-4.39$\pm$0.14&-4.57$\pm$0.24&-4.25$\pm$0.17\\
M$_{W(VI)}$&-5.04$\pm$0.30&-5.09$\pm$0.10&-4.53$\pm$0.13&-4.69$\pm$0.27&-4.37$\pm$0.17\\
\enddata
\end{deluxetable}

\begin{center}
\begin{deluxetable}{lrrrrr}
\tablewidth{5in}
\tablecaption{Galactic Cepheid PLR Zero-points (a) and Slopes (b) \label{tbl-11}}
\tablehead{\colhead{Source\tablenotemark{1}}&\colhead{B07} &  \colhead{B07f}&  \colhead{F01}&\colhead{S04}&\colhead{B03}
}
\startdata
~~~{\bf a}&&&&&\\
V&-4.05$\pm$0.02&-4.03$\pm$0.03&-4.22$\pm$0.02&-4.00$\pm$ 0.10&-4.16$\pm$ 0.22\\
I&-4.78 0.03&&-4.90 0.01&-4.78 0.01&\\
K&-5.71 0.03&-5.67 0.02&&&\\
W$_{VI}$&-5.85 0.03&-5.83 0.03&-5.90 0.01&-5.96 0.04&\\
\\
~~~{\bf b}&&&&&\\
V&-2.43 0.12&-2.44 0.11&-2.76 0.03&-3.09 0.09&-2.69 0.17\\
I&-2.81 0.11&&-2.96 0.02&-3.35 0.08&\\
K&-3.32 0.12&-3.35 0.08&&&\\
W$_{VI}$&-3.37 0.12&-3.30 0.12&-3.26 0.01&-3.75 0.12\\
\enddata
\tablenotetext{1}{B07 = this paper; B07f = no LKH, this paper; F01 = \cite{Freed01};  S04 = \cite{San04}, B03 = \cite{Bar03}. All PLR are parameterized as M = {\bf a} +{\bf b}*(logP-1).}
\end{deluxetable}
\end{center}

\begin{center}
\begin{deluxetable}{lr}
\tablecaption{Reduced Parallax ({\bf RP}) Zero-point Tests of Freedman and Sandage \label{tbl-RPar}}
\tablewidth{0in}
\tablehead{\colhead{Source} &  \colhead{ZP}}
\startdata
F (Eq. 8) & 5.823 $\pm$ 0.036 \\
Freedman   &    5.979\\
Difference     &    +0.156\\ 
\hline
S (Eq. 9) &5.964 $\pm$ 0.042\\
 Sandage  &   5.959\\
 Difference    & -0.005\\
\enddata
\end{deluxetable}
\end{center}

\begin{center}
\begin{deluxetable}{llrrrr}
\tablecaption{Cepheid Parallaxes and Postdictions \label{tbl-parPost}}
\tablewidth{0in}
\tablehead{\colhead{ID}&\colhead{$HST~\pi_{abs}$(mas)} &  \colhead{S06\tablenotemark{a}}&
\colhead{Diff}&  \colhead{F01\tablenotemark{b}}& \colhead{Diff}}
\startdata
\lC&2.01$\pm$0.2&1.74&0.27&1.88&0.13\\
$\zeta$ Gem&2.78 0.18&2.73&0.05&2.64&0.14\\
$\beta$ Dor&3.14 0.16&2.94&0.2&2.84&0.3\\
W Sgr&2.28 0.2&2.34&-0.06&2.19&0.09\\
X Sgr&3 0.18&2.9&0.1&2.69&0.31\\
Y Sgr&2.13 0.29&2.02&0.11&1.84&0.29\\
$\delta$ Cep&3.66 0.15&3.96&-0.3&3.6&0.06\\
FF Aql&2.81 0.18&2.68&0.13&2.39&0.42\\
T Vul&1.9 0.23&1.88&0.02&1.68&0.22\\
RT Aur&2.4 0.19&2.4&0&2.11&0.29\\
unwt'd mean Diff(10)&&&+0.052&&+0.225\\
std err &&& $\pm$0.052 && $\pm$0.039\\
std dev &&& 0.155 && 0.140\\
\enddata
\tablenotetext{a}{Parallax predicted from Equation 9 and Sandage zero-point}
\tablenotetext{b}{Parallax predicted from Equation 8 and Freedman zero-point}
\end{deluxetable}
\end{center}
\begin{center}
\begin{deluxetable}{llrrrrrr}
\tablecaption{LMC PLR Zero-points (a), Slopes (b), and Distance Moduli \label{tbl-14}}
\tablewidth{0in}
\tablehead{\colhead{Source\tablenotemark{1}}&\colhead{Band} &  \colhead{a}&  \colhead{b}&   {a(DC)\tablenotemark{2}}&    {a(RP)\tablenotemark{3}}&\colhead{{\bf DC}(m-M)\tablenotemark{4}}& \colhead{{\bf RP}(m-M)\tablenotemark{4}} }
\startdata
G05&V&14.42$\pm$0.05&-2.78$\pm$0.09&-4.10$\pm$0.04&&18.52 $\pm$0.06&\\
&K&12.78 0.02&-3.26 0.03&-5.70 0.03&&18.48  0.04&\\
&W$_{VI}$&12.65 0.02&-3.37 0.03&-5.86 0.03&&18.51  0.04&\\
\hline
Per04 & K&12.78 0.02&-3.26 0.03&-5.70 0.04& -5.68 0.03 &18.48  0.04&18.45 0.04\\
\hline
OGLE&W$_{VI}$&12.65 0.01&-3.29 0.01&-5.84 0.03&-5.85 0.04&18.49  0.03& 18.50 0.04\\
\enddata
\tablenotetext{1}{G05 = \cite{Gie05}; Per04 = \cite{Per04}; OGLE = private communication G. Tammann. All PLR are parameterized as M = a +b(logP-1).}
\tablenotetext{2}{Zero-points obtained by fitting the data plotted in Figure~\ref{fig-5} but  with  slopes constrained to those from G05, Per04, and OGLE.}
\tablenotetext{3}{Zero-points obtained via reduced parallaxes.}
\tablenotetext{4}{Distance moduli with no metallicity corrections applied.}
\end{deluxetable}
\end{center}

%
%

\clearpage

\clearpage
\clearpage

\clearpage

\clearpage

\clearpage

\clearpage

\clearpage

\clearpage

\end{document}